\documentclass[sigconf]{acmart}

\copyrightyear{2024}
\acmYear{2024}
\setcopyright{rightsretained}
\acmConference[CHI '24]{Proceedings of the CHI Conference on Human Factors in Computing Systems}{May 11--16, 2024}{Honolulu, HI, USA}
\acmBooktitle{Proceedings of the CHI Conference on Human Factors in Computing Systems (CHI '24), May 11--16, 2024, Honolulu, HI, USA}
\acmDOI{10.1145/3613904.3642721}
\acmISBN{979-8-4007-0330-0/24/05}

\AtBeginDocument{%
  \providecommand\BibTeX{{%
    \normalfont B\kern-0.5em{\scshape i\kern-0.25em b}\kern-0.8em\TeX}}}

\usepackage{algorithmic}
\usepackage{textcomp}
\usepackage{xcolor}
\usepackage{hyperref}
\usepackage{cleveref}
\usepackage{subfig}
\usepackage{xspace}
\usepackage{multirow}
\usepackage{makecell}
\usepackage{url}
\usepackage{enumitem}
\usepackage{balance}
\usepackage{graphicx}
\usepackage{color, colortbl}
\definecolor{Gray}{gray}{0.9}

\usepackage{ifthen}
\newboolean{showcomments}
\setboolean{showcomments}{true} % toggle to show or hide comments
\ifthenelse{\boolean{showcomments}}
  {\newcommand{\nb}[2]{
    \fcolorbox{gray}{yellow}{\bfseries\sffamily\scriptsize#1}
    {\sf\small$\blacktriangleright$\textit{#2}$\blacktriangleleft$}
  }
  
  }
  {\newcommand{\nb}[2]{}
  
  }

\newcommand{\translate}{P1\xspace}
\newcommand{\nl}{P2\xspace}
\newcommand{\logging}{P3\xspace}
\newcommand{\pubsub}{P4\xspace}

\newcommand{\eg}{e.g.,\xspace}
\newcommand{\ie}{i.e.,\xspace}

\newcommand{\etc}{\emph{etc.}\xspace}
\newcommand{\etal}{\emph{et al.}\xspace}

\makeatletter
\def\thickhline{%
  \noalign{\ifnum0=`}\fi\hrule \@height \thickarrayrulewidth \futurelet
  \reserved@a\@xthickhline}
\def\@xthickhline{\ifx\reserved@a\thickhline
              \vskip\doublerulesep
              \vskip-\thickarrayrulewidth
             \fi
      \ifnum0=`{\fi}}
\makeatother

\newlength{\thickarrayrulewidth}
\newcounter{RQCounter}
\newcounter{HCounter}
\newcounter{RSCounter}

\usepackage{framed}

\newcommand{\Hyp}[2]{%
\refstepcounter{HCounter} \label{#1}
	\vspace{0.015in}
	\noindent 
	\textbf{H}$_{\arabic{HCounter}}$.~\emph{#2}
	\vspace{0.02 in}
}
\newcommand{\hr}[1]{\textbf{H}$_{\ref{#1}}$}

\newcommand{\mysec}[1]{\vspace{0.08cm} \noindent \textbf{#1.}}

\newcommand{\revisiona}[1]{{{#1}}\xspace}
\newcommand{\revisionb}[1]{{{#1}}\xspace}
\newcommand{\revisionc}[1]{{{#1}}\xspace}

\newcommand{\revision}[1]{{{#1}}\xspace}

\newcommand{\removea}[1]{{}}
\newcommand{\removeb}[1]{{}}
\newcommand{\removec}[1]{{}}
\newcommand{\remove}[1]{{}}

\newcommand{\company}{{Google}\xspace}

\author{Daye Nam}
\affiliation{%
  \institution{Carnegie Mellon University}
  \city{Pittsburgh}
  \country{U.S.A.}
}
\email{dayen@cs.cmu.edu}

\author{Andrew Macvean}
\affiliation{%
  \institution{Google, Inc.}
  \city{Seattle}
  \country{U.S.A.}
}
\email{amacvean@google.com}

\author{Brad Myers}
\affiliation{%
  \institution{Carnegie Mellon University}
  \city{Pittsburgh}
  \country{U.S.A.}
}
\email{bam@cs.cmu.edu}

\author{Bogdan Vasilescu}
\affiliation{%
  \institution{Carnegie Mellon University}
  \city{Pittsburgh}
  \country{U.S.A.}
}
\email{vasilescu@cmu.edu}

\begin{document}

\title{Understanding Documentation Use Through Log Analysis}
\subtitle{An Exploratory Case Study of Four Cloud Services}

\begin{abstract}
Almost no modern software system is written from scratch, and developers are required to effectively learn to use third-party libraries and software services.
Thus, many practitioners and researchers have looked for ways to create effective documentation that supports developers' learning.
However, few efforts have focused on how people actually use the documentation.
In this paper, we report on an exploratory, multi-phase, mixed methods empirical study of documentation page-view logs from four cloud-based industrial services.
By analyzing page-view logs for over 100,000 users, we find diverse patterns of documentation page visits. 
Moreover, we show statistically that which documentation pages people visit often correlates with user characteristics such as past experience with the specific product, on the one hand, and with future adoption of the API on the other hand.
We discuss the implications of these results on documentation design and propose documentation page-view log analysis as a feasible technique for design audits of documentation, from ones written for software developers to ones designed to support end users (\eg Adobe Photoshop).
\end{abstract}




\maketitle

\section{Introduction}
\label{sec:intro}

\revisiona{Almost no modern software system is written from scratch, and many third-party libraries and software services are available to be reused and composed.
Thus, the productivity of programmers in many domains and contexts depends on rapidly searching for relevant information to make decisions about third-party libraries or services~\cite{pano2018factors, myllarniemi2018development}, and learning to use them correctly for their own systems~\cite{robillard2009makes, robillard2011field}.
Practitioners spend a lot of time searching for and digesting relevant API information, \eg 20\% of their time according to \citet{brandt2009two}.
And while many sources are useful, including code examples, question and answer (Q\&A) websites, and expert advice, in obtaining API-relevant information, the official software documentation remains essential~\cite{chen2009empirical, robillard2011field, garousi2015usage}.
}

% Therefore, 
Efforts to improve software documentation span decades,
% Many investigations have been conducted to collect inputs of
with many researchers studying documentation design experts and users to catalogue problems~\cite{chen2009empirical, robillard2009makes, robillard2011field} and recommend best practices~\cite{visconti2004assessing, robillard2009makes, robillard2011field}.
Much documentation now follows such guidelines, and new tools~\cite{Treude.2016, Mehrpour.2019} and ideas~\cite{robillard2017demand} have been proposed to further support developers' information needs based on such studies.

Most of these efforts involve qualitative research methods such as interviews~\cite{nykaza2002programmers, robillard2011field, meng2018application} or lab studies with human participants~\cite{jeong2009improving, horvath2019long, duala2012asking, meng2020optimizing}.
However, while generally highly informative for understanding usability issues during the early design review phase~\cite{DBLP:journals/software/CurtisN95}, such methods capture only what participants say they do, or what they do in a controlled setting.
Moreover, the number of participants that can be observed this way is typically small.

\revisiona{Our research goal is similar to most prior software documentation research---improving the design and usability of documentation. However, our approach is novel and complementary---mining documentation page-view logs at scale.}
Web mining has long been used to analyze people's experience online in more general contexts, \eg user engagement in online news reading~\cite{DBLP:conf/wsdm/LagunL16} or user satisfaction during online shopping~\cite{DBLP:conf/wsdm/SuHLZM18}.
We argue that similar approaches could apply to documentation since, after all, documentation webpages are just another type of webpage.
In other words, any software documentation published on the Web, such as Adobe Photoshop and Autodesk AutoCAD, which comprise numerous documentation webpages and users, could potentially be analyzed using an approach similar to ours.

\begin{figure*}[t]
    \centering
    \includegraphics[width=\linewidth]{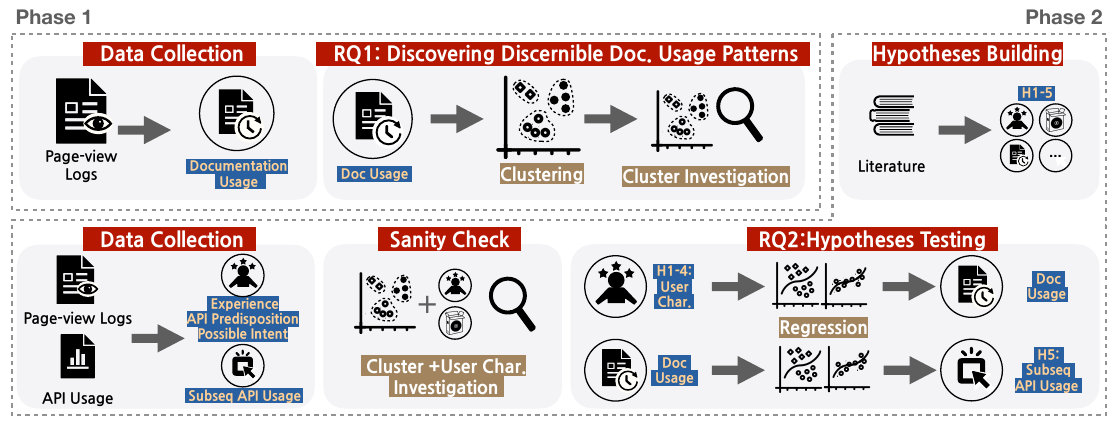}
    \caption{Overview of our data collection and analysis.}
    \Description{This study comprises two phases. Phase one involves two steps: Data Collection and RQ1: Discernible Documentation Usage Patterns. The Data Collection step processes page-view log data into documentation usage. In RQ1: Discernible Documentation Usage Patterns, we conduct a clustering analysis with the documentation usage data and further investigate the clusters. Phase two encompasses Hypotheses Building, Data Collection, Sanity Check, and RQ2: Hypotheses Testing. In this phase, we initiate by formulating hypotheses based on the literature. Subsequently, we gather data for hypothesis testing. In the Sanity Check, we examine clustering results in conjunction with user characteristics data. Finally, in RQ2: Hypotheses Testing, we subject the hypotheses to regression analyses. }
    \label{fig:overview}
\end{figure*}

\revisiona{
We believe that our large-scale log analysis will complement existing documentation review methods, by providing the following additional methodological advantages:
\begin{itemize}
\item \textbf{Allowing more scalable, computational design review}:
    Relying on web analytics to understand documentation usage is considerably less expensive for software providers if they have access to telemetry data for the documentation pages (\eg from self-hosted web servers)---we expect that one quantitative user experience researcher on staff could analyze the page-view logs of hundreds of thousands of users of dozens of APIs or services in a matter of days, if not hours, following our methodology.
    In contrast, qualitative studies tend to focus on one API or service at a time, may require complex participant recruitment, and usually involve orders of magnitude fewer subjects. They also often involve monetary compensation (\eg \citet{duala2012asking} compensated each participant with \$20 for a one-hour programming study), in addition to the researcher team's time for running the studies, and collecting and analyzing the data.
    \item \textbf{Allowing for the discovery of less-studied documentation user groups}: As most of the smaller scale studies require study design prior to the data collection, researchers specify research questions and target participants in advance. For documentation, as it is expensive to conduct these qualitative studies, most of the studies have focused on the professional developers who use the documentation for API learning~\cite{meng2018application, meng2019developers, earle2015user, ko2006exploratory}, the main target usage scenario of software documentation. 
    Large-scale log analysis, on the other hand, does contain the entire user population's data, allowing the discovery of more diverse user groups, including users who use documentation to make API adoption decisions (\eg Product explorers in \Cref{ssec:clustering}), or users who are only concerned with the cost of querying APIs (Financial users in \Cref{ssec:clustering}).
    \item \textbf{Capturing a perspective less prone to response biases:} With qualitative studies where users need to report (\eg survey, interview) or show their behaviors (\eg observation study, lab study), the data can only capture what participants recall or show, which might be different from what actually happens in the wild, \ie response bias~\cite{mccarney2007hawthorne}.
    As qualitative studies often ask participants to focus on ``software documentation regularly used by participants''~\cite{DBLP:conf/doceng/ForwardL02, chen2009empirical, plosch2014value} to help participants recall specifics of their experience, the bias might be even amplified. 
    Log analysis can minimize the response bias, as the telemetry data is automatically collected.
\end{itemize}
}

However, to be clear, traditional non logs-based approaches can be extremely valuable, and we don't advocate replacing them. Instead, we argue that a logs-based analysis like ours could be used as a first pass, to guide the design of more complex (but rich in terms of insights) approaches such as human studies.

To this end, in this paper we report on an exploratory, two-phase, mixed-methods empirical study of documentation page-view logs from over 100,000 users of four popular services \revisionb{of Google}; see Figure~\ref{fig:overview} for an overview.
The documentation page-view logs we had access to were privacy-preserving in a number of ways (section~\ref{subsec:privacy}) and contained only aggregated monthly totals of which specific documentation pages someone visited and how much time they spent on each page, over the course of that month (possibly across multiple sessions). This is likely a common scenario --- many companies and open-source projects can be in a position to instrument their documentation web servers to collect such basic telemetry data; at the same time, it may be undesirable to collect more fine-grained or personally identifiable data for privacy reasons. The research challenge, therefore, is determining whether there is enough signal in this big but shallow data to generate actionable insights for the documentation designers by mining it. 

Overall, our two-phase study argues that the answer is ``yes.''
In Phase I (section~\ref{sec:rq1}) we set out to explore the log data, looking for patterns of page views and trying to explain them \textit{without} knowing who the users are or anything else about them. Given the large size of our sample, we do this using a combination of automatic clustering analysis followed by 
% sampling and 
qualitative explorations 
% of cluster representatives. We 
and show that many page-view clusters are discernible in the log data.

In Phase II (section~\ref{sec:regression}) we set out to formalize and generalize our qualitative observations from earlier. In an effort to understand \emph{why} such discernible patterns exist in the page-view data, we formulate testable hypotheses about the ``average'' characteristics and subsequent behavior of those users, 
based on findings from the literature on general information seeking of developers~\cite{brandt2009two, rao2020analyzing, ko2007information,maalej2013patterns,sillito2008asking,thayer2021theory,DBLP:conf/chi/ZhangHKG20, lawrance2010programmers, freund2015contextualizing} and from small-scale documentation usability studies~\cite{ko2006exploratory, meng2019developers, jeong2009improving, horvath2019long, duala2012asking, meng2020optimizing}.
We then use the fact that all users who make requests to \revisionb{Google} services, or had otherwise registered for accounts on \revisionb{Google} and were browsing the documentation pages while logged in, have persistent (pseudonymized) IDs across the data.
This way, we join the page-view log data with user-level data about their experience with the respective service and the platform overall, and with data about subsequent requests (after the documentation page views) to the service APIs. 
We first revisit the clustering results and check if the hypotheses built based on the general information seeking literature make sense in the documentation usage setting.
We then conduct multiple regression analysis to test the hypotheses formulated in the first phase on this aggregated data, finding multiple sizeable correlations between patterns of documentation page-views, on the one hand, and user-level characteristics and subsequent API use, on the other hand. 
That is, one's level of experience partially explains one's documentation browsing patterns, and one's documentation browsing patterns partially explain one's intent to subsequently use the APIs.

While not intended as an exhaustive exploration of all patterns of documentation page views identifiable for the four \company services in our sample, our study does show that it is feasible to analyze page-view logs at scale to inform documentation design reviews, or to corroborate observations from smaller-scale studies~\cite{costa2009evaluating, meng2018application} or the anecdotal experiences of professional software engineers.
Concretely, we argue (section~\ref{sec:implications}) that even when not knowing anything else about the documentation users, the interaction histories and dwell times that are likely to be contained in the page-view logs can provide actionable information at scale for providers which can help companies decide which documentation pages to redesign, and even to potentially automatically personalize documentation pages in the future, to better align with their users' needs.

\section{Related Work}
\label{sec:literature}

\mysec{Studies on Software Documentation}
% For decades, researchers and practitioners have looked for ways
There have been many investigations to improve software documentation, \eg cataloging problems~\cite{chen2009empirical, robillard2009makes, robillard2011field}, identifying desirable quality attributes~\cite{arthur1992document, dautovic2011automatic}, and recommending best practices~\cite{visconti2004assessing, robillard2009makes, robillard2011field}.
Some of these studies provided concrete insights into what developers need from the documentation.
For example, developers have expressed the need for complete and up-to-date documentation~\cite{aghajani2019software}, because many developers rely on API reference information and code examples~\cite{nykaza2002programmers, meng2018application} when they approach documentation with a problem or task in mind~\cite{meng2018application}.
Developers also asked for a concise overview of the documentation, more rationale, and adequate explanation for code examples~\cite{robillard2009makes, robillard2011field, uddin2015api, meng2018application}.
Researchers have also proposed tools that can assist in more effective usage of documentation, by providing easier access to the documentation contents within developers' workflow~\cite{oney2012codelets, horvath2022understanding, grossman2010chronicle}.

However, most of these studies focused on the documentation artifacts, not the user context.
Still, we know from the literature that the documentation users' needs tend to vary with experience~\cite{ko2011role, earle2015user, maalej2014comprehension,latoza2006maintaining, li2013help, postava2004incorporating}, roles, and learning styles~\cite{li2013help, costa2009evaluating, earle2015user, meng2019developers}. 
For example, \citet{costa2009evaluating} found that documentation users with less experience with the software tended to use more types of documentation than more experienced users, and that tutorials and how-to videos were used by a greater percentage of newer users, and the newer users tended to use tech notes and forums less.
Similarly, in the literature, programmers are sometimes categorized into three personas, which summarize their information seeking and problem solving strategies -- systematic, opportunistic, and pragmatic~\cite{clarke2007end} -- that reportedly also correlate with documentation use~\cite{meng2019developers}.
For example, opportunistic developers tended to use documentation in a task-oriented way, focusing less on the general overview of APIs or the suggestions described in the documentation; in contrast, systematic developers tried to understand how the API works before diving into the details of a task, by systematically searching and regularly consulting documentation provided by the API supplier~\cite{meng2019developers}.
In our study, we provide evidence that these different user characteristics and information needs indeed correlate with their documentation usage, underscoring the importance of accounting for distinct user groups in documentation design.

In terms of research methods, most of the past studies on software documentation relied on interviews~\cite{nykaza2002programmers, robillard2011field, meng2018application}, surveys~\cite{earle2015user, robillard2011field, robillard2009makes, meng2018application}, observation studies~\cite{ko2006exploratory, meng2019developers}, and lab studies~\cite{jeong2009improving, horvath2019long, duala2012asking, meng2020optimizing} that usually involve a small number of participants. 
Our work stands out in that we do not rely on self-reported data, nor lab studies, but rather automatically collected logs of actual documentation page views, thereby offering a complementary approach to studying software documentation and documentation users, based on real-world telemetry data in an industrial context.

\mysec{Studies on Developer Information Foraging}
To learn to use, or reuse, new software frameworks or libraries, developers need a variety of kinds of knowledge~\cite{ko2007information,maalej2013patterns,sillito2008asking,thayer2021theory,DBLP:conf/chi/ZhangHKG20}, so it is important to understand how they search for and acquire information.
There is some prior work on the information seeking strategies of developers, but mostly in general software maintenance~\cite{ko2007information, lawrance2010programmers, freund2015contextualizing} or web search settings~\cite{brandt2009two, rao2020analyzing} rather than learning.
For example, prior work~\cite{brandt2009two, rao2020analyzing} found that developers' web search behaviors vary with their information seeking intent: they visit different types of web pages, use different queries, and overall interact with webpages differently.
In particular, developers were more likely to visit official documentation 
during reminding sessions, versus third-party tutorials
during learning sessions~\cite{brandt2009two}.

More recently, with the advent of large language models, developers have embraced generative models as alternatives to conventional information retrieval from existing sources~\cite{liang2024, fan2023large}. 
Still, researchers have found that the strategies employed by developers to generate necessary information can vary based on factors such as their intent, programming experience, and familiarity with AI tools~\cite{nam2024gilt, zamfirescu2023johnny, ross2023programmer}.

\begin{table*}[t]
\caption{Types of documentation provided for the selected products.}
\label{tab:type}
\small
\resizebox{.95\linewidth}{!}{
\begin{tabular}{ccp{11.0cm}}
\toprule
\textbf{Genre} & \textbf{Type} & \textbf{Description} \\
\midrule
\multirow{2}{*}{Meta} 
& Landing (L)       & Links to core documentation pages.                        \\ %\cline{2-3}
& Marketing (M)    & A brief introduction to a product, incl.\ the benefits, target users, and highlights current customers. \\ \midrule
\multirow{4}{*}{Guide}
& Tutorial (T)      & Walkthroughs for common usage scenarios. \\ %\cline{2-3}
& How-to (H)        & Guidance on completing specific tasks.    \\ %\cline{2-3} % with a product
& Quickstart (Q)    & A quick intro to using the product.                      \\ %\cline{2-3}
& Concept (C)      & Explanations for product- or domain-specific concepts. \\ \midrule  % relevant to a product.
\multirow{2}{*}{Dev}
& Reference (Ref)    & Details about the API elements, including API endpoints and code-level details.     \\ 
& Release note (Rn)  & Specific changes included in a new version.          \\ \midrule
\multirow{3}{*}{Admin}
& Pricing (P)       & Pricing information.                             \\ %\cline{2-3}
& Legal (Lg)         & Legal agreement details. \\ %\cline{2-3}
& Other (O)      & Other resources not included in other types, \eg locations of the servers.\\ 
\bottomrule
\end{tabular}
}
\end{table*}

\mysec{Document Design}
Documentation in fields beyond software development has a richer history~\cite{schriver1997dynamics}. 
Researchers have dedicated their efforts to enhancing document design by delving into audience analysis~\cite{albers2003multidimensional}, refining content based on user feedback~\cite{ko2012lemonaid, bunt2014taggedcomments}, and evaluating the documentation~\cite{mysore2018porta, andersen2012impact}.
However, many of these endeavors were primarily geared towards relatively simpler products like educational brochures or games, which may not fully align with the large-scale software systems with more than hundreds of documentation pages that we analyzed in this work.

\mysec{Studies on Web Usage Mining}
% \subsection{Studies on Web Usage Mining and Personalization}
To improve the usability of web content,
extensive research has been conducted in web usage mining~\cite{DBLP:journals/sigkdd/SrivastavaCDT00}.
By analyzing logs stored in web servers using data mining techniques, it is possible to identify interesting usage patterns, identify different navigational behaviors, and discover potential correlations between Web pages and user groups.
Among the different types of data available in usage logs, page dwell time has been the primary source of understanding users' needs and intentions, along with the search queries~\cite{DBLP:conf/wsdm/KimHWZ14,yi2014beyond,DBLP:conf/aaai/XuZJL08}.
Page dwell time has also been 
% been examined as an indicator of page relevance for web queries, and 
found to correlate with document relevance and user satisfaction~~\cite{fox2005evaluating, DBLP:conf/iui/ClaypoolLWB01, DBLP:conf/sigir/BuscherED09}.
To the best of our knowledge, our work is the first large-scale study of developers' dwell time on documentation pages, showing the feasibility of applying similar approaches in analyzing documentation users' needs.

\section{Dataset}
\label{sec:data}

We started by compiling a dataset of documentation page-view logs for four web-based services of \revisionb{Google}. 

\subsection{Product Selection}

\looseness=-1
\revisionb{Google} provides hundreds of web-based services to a diverse group of users and businesses, and most of the services come with one or more types of APIs, including REST APIs and gRPC APIs, as well as client libraries.
However, since our study is primarily exploratory in nature, we selected only four \company products following a maximum variation sampling strategy~\cite{suri2011purposeful}, to gain an understanding of documentation use from a variety of angles.
Concretely, we diversified our sample in terms of the application domain (machine learning / natural language processing vs. event analytics and management), usage context (operations infrastructure vs. potentially end-user facing), and product size and complexity (ranging from a few API methods to hundreds of API methods offered by the products).
These differences are also reflected in the documentation pages, which vary in their contents across the four products, \eg with more or less marketing materials, how-to guides, pricing information, \etc
All four web-based services we selected are popular, having large user bases.
Specifically, \translate and \nl are machine learning / natural language processing-related products for machine translation and text analysis.
And \logging and \pubsub are operations-related products for managing event streams and log data.

% \subsection{Preprocessing}
\subsection{Documentation Usage Data Preprocessing}
\label{ssec:preprocessing}

For each of the four products, we had access to pseudonymized \textbf{documentation page-view logs}~\cite{liu2007web} for users who visited the documentation from May 1, 2020 to May 31, 2020, UTC, \revisionb{while they were logged into their accounts}.
The page-view log data are collected automatically by the documentation servers and include the specific documentation pages visited by someone, 
as well as the timestamps and dwell times for each visit. 
We use \textit{dwell time} to estimate user engagement with the content, following prior work~\cite{fox2005evaluating,yi2014beyond}.
The data was aggregated at the month level, partially due to the volume of data being analyzed, and also to enhance the pseudonymization of the data for the privacy protection of the users (see \Cref{subsec:privacy} for details).

To reason about more general patterns of documentation use, we further labeled each individual documentation page (URL) in our sample according to its contents into one of 11 possible \textit{types} and four aggregate categories (or documentation \textit{genres}~\cite{earle2015user}) summarized in Table~\ref{tab:type}.
For the first-level labeling we relied on an internal mapping table created by the documentation team, which contains meta information for the different documentation pages, including what we refer to as the \textit{type}.
The second-level labeling reflects our subjective grouping of documentation types into four high-level categories that provide related kinds of information and presentation format; we expect that these are likely to be consulted together given specific tasks and target reader familiarity with the API.
To this end, we followed an open card sorting process involving two authors, one of whom is a domain expert.
There were two documentation types \revisionb{(Other and Release note)} that the two authors did not agree on.
The two authors resolved disagreements \revisionb{by comparing their definitions of genres and the rationale behind the categorization}, until there were no more disagreements.
All official \company documentation pages in our sample were assigned to exactly one of the documentation types and genres listed in Table~\ref{tab:type}, \revisionb{and the documentation of all four products we analyzed included all 11 documentation types.}
\revisionb{The volumes of each documentation type varied, but in general, each product documentation consisted of around 5 pages of Meta genre documentation, around 150 pages of Guide genre documentation, around 300 pages of Dev genre documentation, and around 15 pages of Meta genre documentation.
The documentation of all products followed the same documentation style guide~\cite{google-style}; thus, the contents and styles used for each documentation type are consistent, even across the products.}

Finally, we followed prior work by~\citet{fox2005evaluating} and excluded page-view sessions shorter than 30 seconds, since these are more likely to be noise (\eg a user accidentally clicked the documentation page and then left the page quickly) than meaningful visits.

\subsection{Privacy Protections}
\label{subsec:privacy}

\looseness=-1
\revisionb{As we analyzed the user data of Google, we followed Google's strict privacy and data access policies~\cite{google-safety, google-privacy} 
which ensure appropriate, legal, and ethical access, storing, and analysis of user data. This included, but was not limited to, internal privacy reviews with security and privacy experts, the use of differential privacy processes (more details below), wipeout and data access processes, and more. In addition, the study designs were reviewed by internal research ethics experts, methodological experts, and product experts.
}

We also used numerous privacy protection techniques.
% throughout our work. 
First, all user-level data was pseudonymized before any of the authors had access to it. Pseudonymization maps \revisionb{users' accounts}
% user IDs 
to randomly but persistently generated pseudonymized IDs. As the IDs were randomly generated, they could not be reversed without access to a mapping table, which the authors did not have. 

Additionally, usage data was aggregated (e.g., we looked only at the number of API requests aggregated at the month granularity, not individual API requests), and had Differential Privacy~\cite{wood2018differential} applied. In brief, differential privacy was used to apply sufficient noise to the aggregations such that individual records could not be identified, but the overall shape of the dataset remained meaningful / sufficiently accurate. 
\revision{Using established best practices, and based on guidance with internal privacy experts, we used Epsilon <1.1, (where lower numbers yield higher degrees of privacy protection)}.
This allowed us to analyze trends in user behavior while preserving the privacy of those in the dataset. Later in this paper (e.g., in Figure~\ref{fig:clustering}), we include polar plots of our clusters, but choose only to visualize clusters with over 500 users, as an additional privacy consideration. 
% None of the clusters qualitatively described in this work contain less than 500 people.

\subsection{Limitations \& Threats to Validity}
\label{subsec:threats}
First, a month might not be enough to capture the full process of learning and adopting a complex API, on the one hand, and might not capture the differences in documentation usage patterns that appear in significantly shorter periods, such as patterns in an hour or in a day, on the other hand. 
It is also possible that some users happen to register in the middle of our one-month window, or one may learn an API intermittently over a few months.
However, such an operationalization was necessary to balance data collection complexity, privacy, and analysis scale.
\revisionb{
Given that the size of Google's general user base is very large, and the services we analyzed were already all mature, we believe that our dataset should still capture snapshots of developers at every stage of the learning process, as well as cyclical patterns of use, without the number of newly registered or intermittent users significantly affecting our results.}

\revisionc{The use of a particular month (May 2020) can also be too short to generalize, as the documentation access might change throughout the year, and it could have been influenced by any major event related to the four target products. 
We did our best to choose a month without major events related to the four products we studied, and there was no event for \translate and \nl, but there were two minor feature additions for \logging and two minor beta releases for \pubsub.
However, as we chose to analyze popular products with large active user groups, it is practically not possible to choose a month without any updates.
We believe that selecting four very different products reduces the risk of biasing the results in a meaningful way, especially given that there was no major event that affected all of the four products during that period. 
}

\revision{The documentation and API usage data we used for our analysis can only provide a partial representation of the entire user group's usage.
Since the documentation usage data only include page-view logs of logged-in users, the analysis does not capture the behavior of users who were not logged in, who may behave differently.}
In addition, the aggregated API usage data can only partially represent the outcome of API learning.
For example, while making an API request requires a user to sign into the platform, browsing documentation does not, so not all documentation usage is linked to the corresponding API use.
Multiple developers can make API requests using shared corporate accounts, which can obfuscate the connection between their documentation and API use.

Although the dwell time was logged when the pages were actually accessed, our measurements of time spent on each documentation page are only (over)approximations.
For example, some users may keep a page open without actively consuming it the whole time, while they grab a coffee or read code from their IDE.
As part of our analysis, we applied several heuristics and filters to our data to identify and remove outliers and noise, as described in \cref{ssec:clustering-prep}.

\revisionc{The analyses at the documentation type and genre levels introduce threats to internal validity: the analyses might not capture the possible influence of content and length of individual documentation pages, and other external confounding factors. 
However, the abstraction of data was inevitable due to the number of documentation pages available.
We provide potential ways of introducing additional internal validity control for page-view log analysis in \Cref{sec:implications}.}

While our dataset includes many relevant variables, it certainly does not include all. For example, a user's position or role, their expertise in programming or in the product domain, the specific tasks during which they visited documentation pages, and the actual contents of the documentation pages, are all likely to also correlate with differences in documentation usage but are absent from our data. 
Moreover, we only analyze data for four products of \company, therefore it remains unclear how our findings would generalize.

Finally, the page-view analysis can only be conducted after the documentation has been available for some time, allowing for the accumulation of extensive logs. Therefore, our analysis may not be applicable for documentation writers who need to assess their content pre-release or for documentation related to products with a limited user base.

Thus, we do not expect page-view log analyses like ours to obviate human studies or other more precise research methods. However, we do expect they could be fruitful as a first step or in conjunction with more precise but more costly research methods.

\section{Phase I: Discerning Documentation Use Patterns in Log Data}
\label{sec:rq1}
 
As an initial exploratory investigation to help contextualize our data, we conducted cluster analysis.
This phase was necessary because although we know that developers will use documentation differently, we still know little about how much and in what ways it will differ ``in the wild'' and in our context.
To efficiently explore the large dataset, we first used an automatic clustering analysis to discover discernible documentation usage patterns, and used sampling and qualitative analysis to further investigate the patterns.

\subsection{Data Preparation}
\label{ssec:clustering-prep}

In preparation for clustering, we first aggregated each user's total dwell times (\ie times spent on the different pages) in May 2020 across the 11 documentation page types in Table~\ref{tab:type}. We recorded separate entries for each of our four separate products, if the same user happened to access documentation pages for more than one of the products that month.
We then represented each user's documentation visit profile as a vector of 11 elements, capturing the total times spent across each page type that month. 
Our supplementary material contains a sample of this data and its distribution.

Note that as a precaution before clustering, we filtered out outliers with total dwell times (sum over the 11 page types)
outside of the $[\mu - 3\sigma, \mu+ 3\sigma]$ interval (\ie more than three standard deviations from the mean), as customary. 
In our sample, this corresponds to users who stayed shorter than 1.39 minutes or longer than 961.91 minutes in total across all documentation pages of each of our four products \revision{during the month} (in May 2020).
In addition, as the distributions of dwell times we observed tend to be right-skewed, we log-transformed all positive values. This is a common transformation~\cite{romesburg2004cluster} when the data vary a lot on the relative scale, as in our case --- spending one more minute on a page is arguably much more noticeable for a 3-minute dwell time than a 10-hour dwell time.

\subsection{Methodology}
\label{ssec:clustering-method}

Out of many clustering approaches available, we adopted a protocol proposed by \citet{zhao2016discovering}, which is particularly well suited for large datasets.
A common challenge with standard clustering methods is determining the appropriate number of natural clusters. Typically, one either chooses the number of clusters a priori, or applies techniques to automatically determine the ``optimal'' number of clusters. The former scenario is not applicable in our case (we do not have any empirical basis to expect a particular number of clusters), while traditional techniques to select the number of clusters automatically tend to be slow for large datasets like ours.
The key innovation in the protocol by \citet{zhao2016discovering} is combining two standard clustering techniques: first using a fast clustering method (k-means) to reduce the dimensionality of the clustering problem, and then applying a second clustering method (MeanShift) that automatically determines the number of clusters.
This is computationally effective, as the second method only runs on the centroids generated by the first (k-means).
To select the number of clusters as input for the first (fast) method, one typically chooses a significantly larger number than the plausible number of natural clusters, expecting that the second method will merge closely located centroids eventually to match the natural clusters.
In determining the quality of the clustering results from the different parameters used, we used the following clustering performance score, as per Zhao~\etal.
\vspace{-.3cm}

\begin{align}
\label{eq:cp}
\vspace{-.7cm}
cp = 0.3 * E + 0.23 * D + 0.23 * \frac{k-m}{k} + 0.23 * \frac{N-n}{N}
\end{align}

In this score, the first and the second factors are used to reward the clustering performance using two well-known metrics: Shannon's entropy (E) and Dunn's index (D).
The probability used for calculating Shannon’s entropy score is the normalized number of users in each cluster. 
Thus, entropy assigns a high value to clustering results that have a uniform distribution of users across clusters.
Dunn's index measures the compactness and separation of the clusters, by calculating the ratio of the smallest distance between observations not in the same cluster to the largest intra-cluster distance.
The third and fourth factors are to penalize clustering results that are too naive or complex.
The third penalizes the results that are too complex, that do not improve over the naive k-means results, where $m$ (the number of clusters after MeanShift clustering) is close to $k$ (the number of target k-means cluster that is significantly larger than the number of natural clusters).
The fourth factor penalizes results that one big cluster contains most of the users in the dataset, where N is the number of total product users and n is the number of users in the biggest cluster.

After trying several values for $k$ and $eps$ in the equation~\ref{eq:cp}, we obtained the highest $cp$ score, 0.50, which is comparable to other works~\cite{zhao2016discovering, DBLP:journals/tois/TianZP23},
with E = 0.71, D = 0.15, m = 316, N = 94096, n = 9789.
This result was obtained for $k=400, eps=1.25$, resulting in 320 clusters.
Most clusters consist of around 300 users and there are 18 clusters consisting of more than 1,000 users.

\begin{figure*}[htp!]
  \centering
  \includegraphics[width=0.9\linewidth]{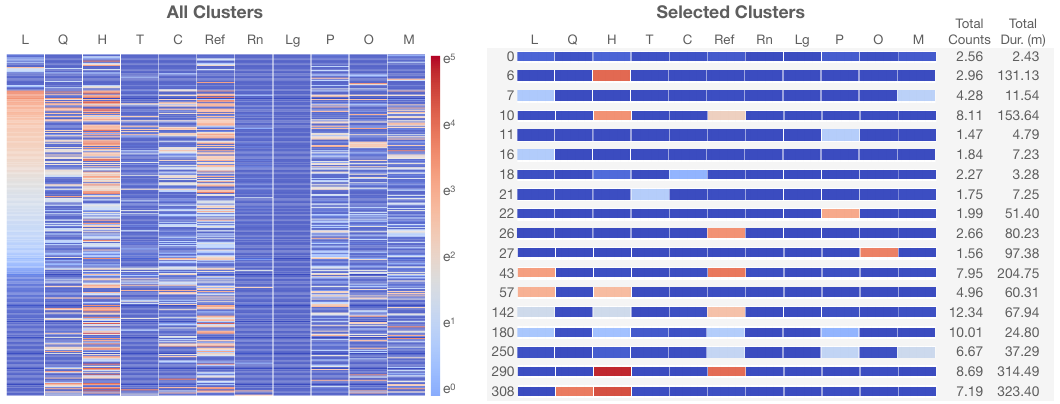}
  \caption{The heatmap of centroids of the 320 clusters (left), and a subset of them highlighted (right). Each row represents the documentation usage of each cluster (see Table~\ref{tab:type} for the documentation type codes). The color indicates the dwell time in minutes\revision{, with the intensity encoded in $e^n$ of time}. \revision{The average total counts (\# of documentation pages visited in May) and the average total dwell time (sum of dwell time on 11 documentation types) are also shown for the selected clusters (right) to help with interpretation, and the rows are sorted by the average total dwell time. For example, users of Cluster 18 (2nd row from the selected clusters) spent 3.28 minutes on average on the product documentation among 2.27 page visits on average, and spent $\approx e^1=2.7$ minutes on Concept type documentation. }}
  \Description{Left: A color-coded heatmap displaying centroids of 320 clusters representing documentation usage. Each row corresponds to a cluster type (L: Landing, Q: Quickstart, H: How-to, T: Tutorial, C: Concept, Ref: Reference, Rn: Release note, Lg: Legal, P: Pricing, O: Other, M: Marketing). The heatmap uses color to indicate dwell time in minutes, with varying colors representing different levels of engagement. The color scale ranges from \math{e^0} to \math{e^5}, indicating increasing levels of dwell time. Right: A subset of the heatmap enlarged for visibility, reported with their total counts and total dwell time in minutes.}
  \label{fig:heatmap}
\end{figure*}

 \subsection{Resulting Clusters}
 \label{ssec:clustering}

Figure~\ref{fig:heatmap} provides a summary of the resulting clusters, illustrating a wide array of documentation usage patterns within the 320 clusters. These patterns are evident through the variations in page views across the 11 documentation types. To comprehensively investigate these distinct patterns, after excluding small clusters with fewer than 100 users, we conducted open coding. The primary author assigned codes to the clusters, refined them iteratively, and subjected the emerged categories to a thorough review by the entire team of authors. In light of space limitations, we are able to present only four codes that were frequently assigned to clusters, along with representative examples of these clusters for reference.

\mysec{Product explorers (Clusters 11, 16, 21)}
The time this group spent on documentation is not seemingly enough to digest the information in the documentation, and would not help one to actually use a product.
Furthermore, considering that the clustering was done with a month-full of user logs, visiting only one type of documentation for few times is not likely a usage pattern of an actual product user.
Cluster 16, for instance, only visits one specific type of documentation, landing, a few times, and spent only a short time on the documentation (less than 10 minutes) on average. 

\mysec{Documentation Explorers (Clusters 7, 18, 180)}
Users in Cluster 180 were similar to product explorers in the sense that they only stayed for a relatively short time (less than 30 minutes over the month), but different in that they visited more documentation pages of multiple types.
We infer that these users might be new to the documentation and might be exploring it to see the available information.
For example, we speculate that they might have visited Landing documentation by searching for the product name in search engines, checked the prices from the Pricing page to see if they can adopt the product, and looked around Reference documentation to see the features available.

\mysec{Task-oriented users (Clusters 6, 26, 27)}
Users of Cluster 6 showed more distinct behaviors.
Although they only visited one specific type of documentation a few times, like product explorers, they stayed on the documentation pages much longer (on average, 131.13 minutes over the month).
Based on the amount of time spent on documentation pages, we can infer that these users spent enough time to find what they were looking for from the information, and to fully digest it. 
From the number of visits, the users did not seem to explore the documentation, but stayed on few specific (or single) pages that they were interested in; this indicates that they were only interested in some of the product features, rather than an overall understanding of the product.

\mysec{Versatile users (Clusters 43, 290, 308)}
These users visited multiple types of documentation pages and stayed long enough time on each of the type.
For example, users in Cluster 290 
seemed to be interested in the specific tasks described in How-to guide pages since they spent enough time on them, and perhaps visited Reference documentation from time to time when they needed more low-level information on the API calls and parameters.

We also discovered many other interesting documentation usage patterns, such as \textbf{Financial users} (Cluster 22), who stayed in Pricing documentation which only contains pricing information for an hour, and \textbf{Server engineering users} (Cluster 27), who almost exclusively visited Other documentation which provided resource-relevant information like locations of the servers.

\section{Phase II: Factors Associated with Documentation Use}
\label{sec:regression}

With the exploration of clustering analysis results, we were able to \revisiona{discover} various usage patterns, including those that \revisiona{were not actively discussed} in the existing literature\revisiona{~\cite{meng2018application, meng2019developers, earle2015user, ko2006exploratory}}, like product explorers or documentation explorers. 
However, we could only speculate about the intention and background of the users behind those diverse documentation usage patterns.
Thus, we now bring together our informal observations from Part I with the literature on general information seeking and small-scale documentation studies, to derive and test hypotheses explaining the different usage patterns based on user characteristics.

\subsection{Hypotheses Building}
\label{subsec:hypotheses-building}

Given the absence of established theoretical frameworks elucidating documentation usage behaviors, we have chosen factors that might be associated with the developers' documentation usage, informed by the prior work on developers' general information seeking in web search or software maintenance settings, and observations from the small-scale documentation studies~\cite{freund2015contextualizing, meng2019developers, clarke2007end, li2013help, costa2009evaluating}.

\vspace{0.3cm}
\mysec{Experience}
Many studies have shown that the information seeking strategies of developers vary by their experience levels~\cite{ko2011role, earle2015user, maalej2014comprehension,latoza2006maintaining, li2013help, postava2004incorporating}.
For example, \citet{costa2009evaluating} found that documentation users with less experience with the software tended to use more types of documentation than more experienced users, and that tutorials and how-to videos were used by a greater percentage of newer users, and the newer users tended to use tech notes and forums less.
Thus, we hypothesize,

\Hyp{hyp:experience}{High experience levels are positively associated with accessing documentation covering implementation details (Dev genre), whereas lower experience levels are positively associated with accessing documentation covering an overview of the products (Meta genre).}

\vspace{0.3cm}
\mysec{Product Type}
Differences in typical usage contexts of the products, such as project complexity and task categories, also influence developers general information seeking~\cite{freund2015contextualizing, montesi2008classifying, earle2015user}.
% could also explain the different usages of documentation
Within our dataset, \translate and \nl are application APIs whereas \logging and \pubsub are operations-related products for managing event streams and log data, and we expected to see different characteristics will come with different documentation usages, and we anticipate that these distinct characteristics will be associated with different documentation usage patterns. 
For instance, users of infrastructural APIs are more likely to be engaged in the maintenance of large-scale software projects, which implies a greater interest in system-level products and in system-level quality attributes. Conversely, application APIs are commonly adopted by smaller projects where the applications themselves serve as core components. 
Consequently, we expect that users of documentation for different products will tailor their utilization accordingly. Thus, we hypothesize:

\Hyp{hyp:product}{Documentation usage of application APIs (\translate, \nl) differs from that of  large-scale infrastructural APIs (\logging, \pubsub).}

% \noindent To analyze the differences in documentation usage patterns, we recorded what each documentation usage data point was for.

\vspace{0.3cm}
\mysec{Documentation Type Predisposition}
Previous research has found that developers adopt different work styles, motivations, and characteristics, and they solve programming tasks differently~\cite{clarke2007end}, and human studies with documentation usage also reported  similar findings~\cite{meng2019developers}. 
The work styles of developers are less liable to change over time as opposed to levels of expertise, educational background, etc.
Thus, we hypothesize that we can see similar patterns in the page-view logs, that developers will stick to documentation that suits their general information foraging strategy formed by their needs and preferences, without changing their documentation usage behavior much over time.

\Hyp{hyp:strategy}{Users tend to use the same documentation type over time.}

\vspace{0.3cm}
\mysec{Possible Intent}
Prior work~\cite{brandt2009two, rao2020analyzing} found that developers' web search behaviors vary with their information seeking intent: they visit different types of web pages, use different queries, and overall interact with webpages differently.
In particular, developers were more likely to visit official software documentation
during reminding sessions and third-party tutorials during learning sessions.
Developers also tend to spend tens of minutes with learning intent, but only tens of seconds to remind. Times spent in between the two extremes were mostly with clarification intent.
We posit that a similar behavioral pattern can be identified within documentation-based information seeking, wherein users invest substantial time per visit when their objective is to grasp complex concepts or protocols. Conversely, they allocate less time when verifying straightforward facts or utilizing documentation as an external memory aid~\cite{ko2006exploratory}.
Thus, we hypothesize:

\Hyp{hyp:intention}{Users who exhibit extended average page dwell times are more inclined to access documentation that offers tutorials (Guide genre), whereas users with shorter average page dwell times are more likely to access documentation providing straightforward factual information (Admin genre).}

\vspace{0.3cm}
\mysec{Subsequent API Use}
From multiple empirical studies, developers have reported that the quality of documentation is a highly influential factor in API selection process~\cite{DBLP:conf/sigsoft/VargasATBG20}, and failure in effective information seeking within documentation leads them to give up on using the APIs~\cite{10.1109/vlhcc.2016.7739689}.
Developers specifically reported that they examine the documentation up-front to determine ``if there are good examples or tutorials that clearly explain how to use the library''~\cite{DBLP:conf/sigsoft/VargasATBG20} before they decide to adopt a library, showing the need of onboarding materials.
Thus, we expect that:

\Hyp{hyp:future}{Accessing documentation pages providing technical information for newcomers (guide genre) is positively associated with subsequent API calls by the same users.}

\subsection{Data Preparation}

We collected \textbf{pseudonymized user-level data} and \textbf{API usage data} to extract such factors of \company's documentation users, and test whether the hypotheses in the previous section are supported by the developers' documentation usage data at scale.

\mysec{Experience}
\noindent To investigate the effect of experience levels in documentation usage, we measure the documentation users' experience level using two variables: \textit{overall platform experience} and \textit{specific product experience}.
We define the experience with the platform as the user account age, \ie years passed since signing the platform terms and conditions\footnote{For users who have never signed the terms and conditions for API usage, we assigned a value of 0.}.  
We define the experience with a specific product as the total number of successful API requests made to that API over the previous three months (February, March, April 2020).

\mysec{Product Type}
\noindent To analyze the differences in documentation usage patterns, we recorded what each documentation usage data point was for.

\mysec{Documentation Type Predisposition}
\noindent As a proxy for one's possible predisposition for certain documentation types, we recorded the user's documentation \textit{page views in the previous three months} (February, March, and April 2020).

\begin{figure*}[t]
    \centering
    \includegraphics[width=\linewidth]{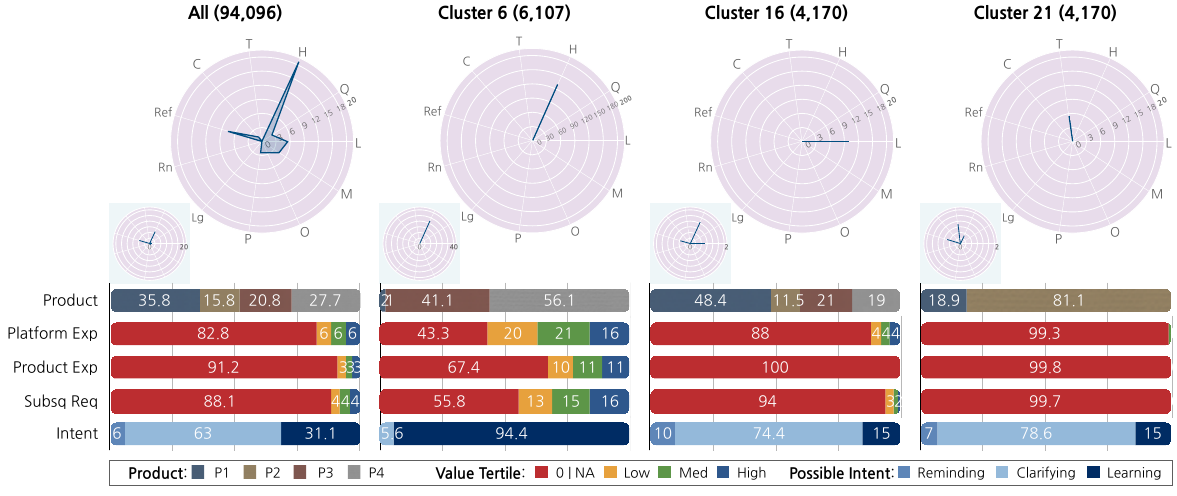}
    \caption{Highlights of the clustering analysis. Each polar plot displays the average time spent on each type of documentation (see Table~\ref{tab:type} for the documentation type codes). The small polar plots show the average dwell time in the previous three months. Note that the ranges of the axes of the plots vary. Bar charts below the polar plots show the proportions (\%) of each group in the cluster. \revision{For example, the charts of Cluster 21 can be interpreted as ``In cluster 21, users without platform and product experience predominantly used Tutorial documentation ($\approx$ 6 minutes) of \nl (81.1\%) and \translate (18.9\%), mostly for clarification purposes, without subsequent API requests.''}
    }
    \Description{Polar plots of average dwell time of selected example clusters accompanied with bar charts summarizing the user distributions. The first set shows the average usage of the entire users in the dataset, showing the average usage of 20 minutes on How-to guides, more than 7 minutes of reference documentation, and less than 5 minutes on landing, marketing, and pricing documentation. The products of the users were relatively equally distributed. More than 80\% of the users had no platform and product experience more than 80\% did not make any subsequent requests on average. More than 60\% had clarifying intention and 30\% had learning intention. For clustering 6, the users spend around 130 minutes on How-to guide documentation. Most of the users in this cluster used \pubsub or \logging. More than 50\% of the users had used the \company platform in the past, and around 40\% of the users had used the product in the past. Around 50\% of the users made subsequent API requests, and almost 95\% of the users had learning intent. Cluster 16 users spent around 10 minutes on landing pages. The products that users used were relatively equally distributed. Only around 10\% of the users had experience with the \company platform, and there was no user with product experience. Only 6\% of the users made subsequent API requests, and 75\% of the users were with clarifying intentions. For cluster 21, users spent less than 6 minutes on tutorial documentation on average. Most of the users used \translate or \nl, and almost no users had experience with the platform, or product. Also, almost no user made subsequent API calls, and around 80\% of the users were with clarifying intent. }
    \label{fig:clustering}
\end{figure*}

\mysec{Possible Intent}
\noindent As a proxy for possible user intent when accessing the documentation, we recorded the \textit{average per-page dwell time}, by dividing the total dwell time in May by the total number of documentation pages a user had visited.\footnote{A more direct comparison to \citet{brandt2009two} would require per-session dwell times, which we did not have access to, hence this approximation.}
We further grouped the data into three bins---less than 1 minute, between 1 minute and less than 10 minutes, and more than 10 minutes---to loosely correspond to the categories of intent (reminding, clarifying, and learning) identified by \citet{brandt2009two}. The ``more than 10 minutes'' group most directly maps to a learning intent, while the other two groups possibly overlap with both reminding and clarifying.

\mysec{Subsequent API Use}
\noindent We mined the \company-collected API usage data (telemetry data) from June, July, and August 2020 corresponding to the subset of people in the aforementioned May-2020 documentation page-view log dataset, who also made subsequent API requests using the web-based services. This was possible because the pseudonymization strategy has random but persistent IDs that are consistent across \revisionb{documentation and API usage data}.
Specifically, we extracted the \textit{number of successful API requests} made by each user 
(\ie with 2XX return codes).

% \subsection{Cluster Exploration}
\subsection{Sanity Test with Cluster Exploration}
\label{subsec:sanity-test}

Before we formally test our hypotheses, we checked whether the hypotheses derived based on the general information-seeking literature apply at all to developers' documentation usage patterns observed in our data, by checking different clusters' user distributions.
To help with our exploration,
we first visualized each cluster's average dwell time, and
discretized the numerical variables into four groups for each user factor, based on percentiles: \path{0|NA} (factor=0), \path{Low} (0-33\%), \path{Medium} (34-66\%), \path{High} (67-100\%).
Figure \ref{fig:clustering} shows the visualizations of the entire dataset, and three example clusters due to the space limit. 
We have included visualizations of other large clusters with over 500 users in our supplementary materials.\footnote{To protect privacy (see \cref{subsec:privacy}), we have not included visualizations of the remaining clusters. However, we note that these large clusters account for 77\% of the total users in our dataset.}
Using the visualizations, we selected clusters with different distributions for each factor we hypothesized would influence documentation usage.
We then compared their documentation usage patterns to check if the factors were related to variations in these patterns.

\hr{hyp:experience}\textbf{ (Experience):} 
Comparing clusters with a lot of experienced users (\eg Cluster 6, Cluster 26, Cluster 27) and clusters mostly with new users (\eg Cluster 16, Cluster 21, Cluster 22), the dwell time of the latter was relatively shorter compared to the former.
We also found that most of the clusters with more experienced users spend time on the documentation that describes lower-level details, such as Reference or How-to guide documentation, without needing to visit introductory documentation like landing or marketing pages.
On the other hand, clusters with new users showed diverse documentation usage patterns, which might be because they browse the documentation while considering adopting the APIs while still being relatively unfamiliar with the products, instead of trying to learn to use the products.

\hr{hyp:product}\textbf{ (Product type):}
We observed that documentation usage for \translate and \nl, on the one hand, and \logging and \pubsub, on the other hand, is internally similar in different clusters --- many users of the pairs ended up clustered together (\eg  Cluster 21 and Cluster 11 for \translate and \nl, and Cluster 6 and Cluster 10 for \logging and \pubsub).
Clusters with a lot of \logging and \pubsub users visited How-to guide documentation, which might be due to their typical high project complexity requiring system-level configurations of multiple products in \company platform.
In addition, we observed that clusters with the majority of users using application APIs show longer pricing documentation usage, whereas clusters of infrastructural API users show almost zero usage of pricing documentation.
This could be explained by the usage context of the products: Infrastructural API users maintaining large software systems are also often employees of large corporations, with accounting and legal teams taking care of administrative tasks, removing the need to visit Pricing or Legal documentation,
whereas application APIs are often used by smaller companies or individual projects whose developers are more likely to be responsible for administrative tasks.

\hr{hyp:strategy}\textbf{ (Documentation type predisposition):}
We observed a consistent trend where clusters of users who spent an extended amount of time on specific types of documentation in May also exhibited a prolonged engagement with the same documentation in previous months. 
For instance, consider Cluster 6 \revision{(task-oriented users)}, whose members demonstrated a substantial dwell time on How-to documentation in May; they also ranked second in terms of How-to documentation usage in previous months, among the clusters we analyzed.
Similarly, Cluster 22 \revision{(financial users)}, which had the longest dwell time of Pricing documentation in May, consistently showed the longest dwell time for Pricing documentation in preceding months. 
Furthermore, even among clusters with lighter documentation usage, we noticed a parallel pattern: the dwell times from previous months mirrored the patterns observed in May.

\hr{hyp:intention}\textbf{ (Possible intent):}
Comparing clusters with a lot of users with ``reminding'' or ``clarifying'' intention (\eg Cluster 0, Cluster 16, Cluster 18) with clusters with mostly ``learning'' intention (\eg Cluster 6, Cluster 26, Cluster 27), we observe that the former users spent much less time on the documentation pages on average, and also focused on documentation types like marketing and landing pages, which often provide an overview and administrative facts of the APIs, consistent with the ``reminding'' and ``clarifying'' intent reported by \citet{brandt2009two}.
In contrast, clusters with a lot of ``learning'' users visited documentation that provides more detailed guidance on how to use the products, like How-to documentation.

\hr{hyp:future}\textbf{ (Subsequent API use):}
Comparing the clusters of users who made no or low subsequent API calls (\eg Cluster 0, Cluster 16, Cluster 22) with the clusters of users who made subsequent API calls (\eg Cluster 6, Cluster 26, Cluster 27), the latter
had spent longer overall browsing documentation pages, and had spent most of their time on How-to guide and Reference pages as opposed to marketing pages, which could indicate that many had already decided to adopt the API.
We also observed that the degree of such association may vary with the product.
For example, compared to the users in Cluster 7 \revision{(documentation explorers)} who visited Landing and Marketing documentation and had similar average dwell times, far more users in Cluster 16 \revision{(product explorers)} actually made calls to the API in the subsequent months.
This might be explained by their usage context: the product proportions were relatively equal in Cluster 16, but most of the Cluster 7 users visited only \translate documentation.
Thus, we expect that users will need different types of information depending on their usage context, and thus the usefulness of documentation types may also vary.

\begin{figure*}[p]
    \centering
    \includegraphics[width=0.63\linewidth]{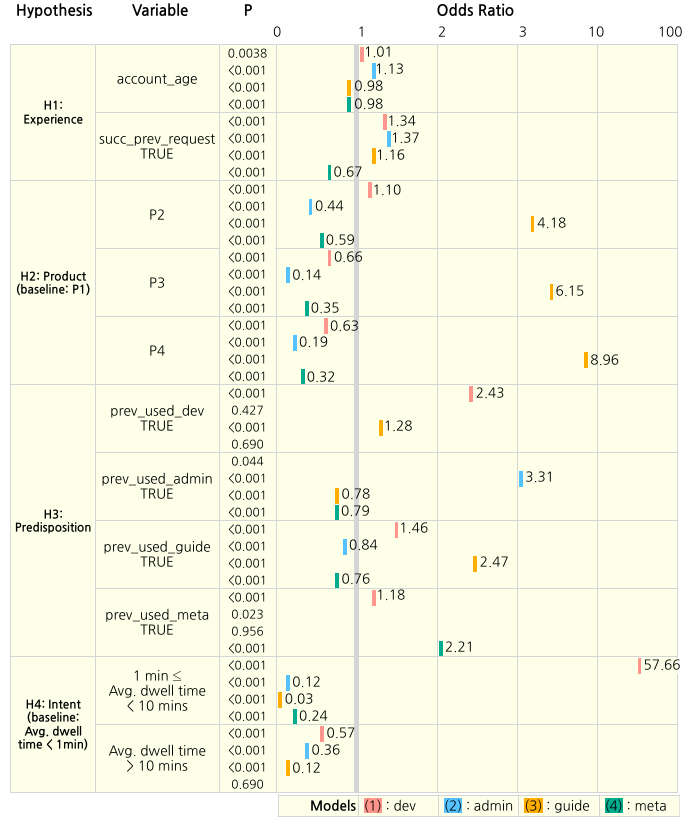}\\
    \includegraphics[width=0.63\linewidth]{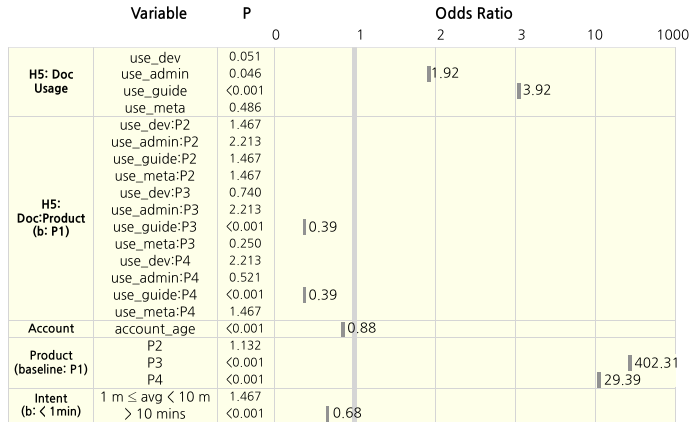}
    \caption{\textit{Top}: Estimated odds ratios from the regression modeling $\text{\textit{dwell time}} > 0$ for our four documentation genres. 
    \revision{For example, the odds of accessing Dev type documentation (pink) are 1.01 times higher among users with one extra year of platform experience.}
    \textit{Bottom}: Estimated odds ratios from the regression modeling $\text{\textit{subsequent requests}} > 0$.
    Variables without statistically significant coefficients (adjusted $p \geq 0.01$) are omitted.
    }
    \Description{Top: Scatter plots of odd ratios from the regression modeling dwell time > 0, for the 4 hypotheses. Most of the variables show statistically significant results, where the typical odds ratio ranges from 0.03 to 2. Bottom: Scatter plots of odd ratios from the regression modeling subsequent requests > 0, for the \hr{hyp:future}. For the doc usage variable, only use\_admin and use\_guide show statistically significant differences, with 1.92 and 3.92 as the odds ratios. For the interaction terms, use_guide:P3, and use_guide:P4 show statistically significant results with odds ratios being both 0.39.}
    \label{fig:rq2-logistic}
\end{figure*}

\subsection{Regression Analysis}
Next, we formally test the hypotheses above on our entire sample.
First, we use multiple regression to test how much the various user-level characteristics we hypothesized about in \hr{hyp:experience}-\hr{hyp:intention} can explain people's logged documentation visits to pages in each of our four genres (recall Table~\ref{tab:type}). 
Second, we test \hr{hyp:future},
\ie to what extent developers' documentation visits in each of our four genres can explain their subsequent API use, again using multiple regression.

We start by estimating four logistic regression models, one for each documentation genre; see model specification in the supplementary material.  
In each model, the dependent variable is a boolean variable ``$\text{\textit{dwell time}}>0$'' indicating whether or not a user in our sample accessed documentation pages of that particular genre.\footnote{See the supplementary material for consistent results for complementary count-based, linear regression models that further investigate the time spent on the different pages.} 
In addition, each model includes explanatory variables corresponding to 
\hr{hyp:experience} (overall platform experience, specific product experience),
\hr{hyp:product} ($\text{product}$),
\hr{hyp:strategy} (documentation use in the previous three months), %Documentation Type Predisposition.
and \hr{hyp:intention} ($\text{average page dwell time}$);
see section~\ref{ssec:preprocessing} for definitions. All models include all variables.
By jointly estimating the different $\beta$ coefficients, this model allows us to estimate the strength of the association between each explanatory variable and the likelihood that users access documentation pages from each genre, \textit{independently of the other variables included in the model}. 
Then, the $p$-value of, say, the estimated $\beta_{1}$ coefficient allows us to test \hr{hyp:experience}, \ie whether there is a correlation between platform experience and the likelihood of accessing documentation genres being modeled.
Similarly, we could test for correlations between platform experience and likelihood of accessing documentation pages from the other three genres with the other three models.\footnote{Note that our research hypotheses in section~\ref{subsec:hypotheses-building} are not all equally broad, \ie they don't all cover all documentation genres or even the same documentation genres. Our choice to model each documentation genre separately is flexible enough to allow us to draw conclusions about all hypotheses, including the broader ones, by qualitatively comparing results from the relevant models. For example, we can reason about a particular estimated coefficient $\beta$ being statistically significant in multiple models corresponding to multiple documentation genres.}
                  
To test \hr{hyp:future}
we use a similar strategy, estimating one logistic regression model with a boolean-dependent variable 

``\textit{subsequent requests} $>$ 0''.
We restrict this analysis to the subset of users who have not made any API requests in the past months (more likely to be new users), since we expect the results to be more actionable for this subset in terms of growing the API user base. 
We include all the same independent variables as before (the ones not directly tied to the hypotheses act as controls), except specific\_product\_experience which is by definition null for these users. 
We also include an interaction with product to test the effect of differences in products.

Overall, we took several steps to increase the robustness of our estimated regression results. First, we removed outliers (\ie observations more than 3 standard deviations beyond the mean) for highly skewed count variables and applied log-transformations to improve heteroskedasticity. Second, we checked for multicollinearity using the Variation Influence Factor (VIF) and only kept variables having VIF lower than 2.5, following \citet{johnston2018confounding}. Third, since we estimate multiple models, each with multiple variables, thus increasing the risk of Type I errors, we conservatively adjusted all $p$-values using Holm's correction procedure~\cite{holm1979simple}. Furthermore, we only considered model coefficients worthy of discussion if the adjusted $p$-values were statistically significant at 0.01 level instead of the more common 0.05.

\subsection{Results}
\label{ssec:regression_results}

Figure~\ref{fig:rq2-logistic}-top summarizes the documentation-access logistic regression results across the four models we estimated (one per genre) to test \hr{hyp:experience}-\hr{hyp:intention}. 
We present our results in terms of odds ratios (OR) instead of regression coefficients to ease interpretation.
All four models are plausible, with Nagelkerke~\cite{nagelkerke1991note} pseudo $R^2$ values (deviance explained) of 74\% for dev, 16\% for admin, 44\% for guide, and 55\% for the meta documentation genre.
Similarly, Figure~\ref{fig:rq2-logistic}-bottom summarizes the subsequent-usage logistic regression model testing \hr{hyp:future}.
% and \hr{hyp:interaction}.
The relatively high explanatory power of the models indicates that at least some of the patterns of documentation usage align with user characteristics and API usage behaviors.

\mysec{\hr{hyp:experience} (Experience): supported}
Results from the dev and meta-genre models are consistent with the hypothesis. For example, the odds of accessing reference documentation and other dev pages are 1.34 times higher among people with prior experience with the products (product experience), \ie those who made successful API requests in the past, compared to those without, and the odds of accessing such pages are 1.01 times as high among users with one extra year of platform experience.
Similarly, the odds of accessing marketing and other meta documentation are lower ($\text{OR}=0.67$) among people with prior experience with the products (product experience), and the odds of accessing such pages are 0.98 times as high among users with one extra year of platform experience.

Interestingly, the results from the admin-genre model align more with the documentation genres covering implementation details than meta: the odds of accessing pricing, legal, and other admin documentation are also higher (1.37 times) among people with prior experience with the products compared to those without.
This could indicate that the information in admin documentation is not only needed once, when people make API adoption decisions, but rather is consistently needed throughout their use of the API.

\mysec{\hr{hyp:product} (Product type): supported}
All four models support the hypothesis: taking \translate as the reference, the magnitude of differences between \translate and \nl is consistently smaller than either \translate and \logging or \translate and \pubsub; \ie the documentation page visits of large-scale infrastructural products tends to differ starkly from that of application products.
Taking the dev-genre model as an example, the odds of accessing the documentation pages are only 1.1 times higher among visitors to \nl documentation compared to \translate, but 0.66 and 0.63 times as high among visitors to \logging and \pubsub compared to \translate. 
% respectively.

\mysec{\hr{hyp:strategy} (Documentation type predisposition): supported}
All models show strong effects of documentation type consistency: \revisionc{there are correlations between the past and the future access to some types of documentation.}
For instance, in the admin-genre model the odds of accessing admin-genre documentation pages are 3.31 times higher among people who had also accessed such pages in the past three months compared to people who had not.
\revisionc{As many different pages of documentation are included in each type of documentation, and the analysis is done at a month-level,
this result does not provide conclusive evidence of the users' preference for the contents or structure of documentation pages. However, it still suggests that the documentation users have types of documentation they are more familiar with, and can access them repeatedly.  }

\mysec{\hr{hyp:intention} (Possible intent): only partially supported}
The results for this hypothesis are mixed. 
On the one hand, the dev-genre model reveals a clear difference between people with long and short average per-page dwell times, as hypothesized: the odds of accessing reference documentation and other dev pages are 0.57 times lower among people with average dwell times greater than 10 minutes compared to those with average dwell times less than a minute. 
The model also reveals that the odds of accessing dev-genre documentation are greatest (57 times higher) among people with average dwell times between one and 10 minutes. %compared to those with average dwell times less than a minute. 
Similarly, the models for admin- and meta-genre pages, which include marketing and pricing, are generally supporting the hypothesis.

In contrast, the model for guide-genre documentation points to the opposite finding than hypothesized when comparing to people with average dwell times less than a minute (the group with the shortest dwell times, set as the baseline in our models): the odds of accessing tutorials, how-to documentation, and the like are lower, not higher, among both people with average dwell times between one and 10 minutes as well as people with average dwell times greater than 10 minutes, compared to those with average dwell times less than a minute. 

One potential explanation is that many users might use the guide documentation as a cheat-sheet, from where they copy and paste various API boilerplate~\cite{nam2019marble} or usage examples.
Although guide documentation was originally intended to introduce and explain products to relatively inexperienced users, it appears to be widely used by users with diverse intentions.

\mysec{\hr{hyp:future} (Subsequent API use): supported}
\label{subsubsec:h6}
The model reveals a strong correlation between accessing guide-genre documentation pages and subsequent API calls: the odds of making successful API calls in the subsequent three months are 3.92 times higher among people who visited guide documentation
compared to those who did not.

Modeling interactions revealed that the strength of the association between visiting guide documentation and making subsequent API requests is weaker for \logging and \pubsub users relative to \translate, while the interaction is not statistically significant for \nl users relative to \translate.
That is, as above, we see consistent differences between the two large-scale infrastructural APIs and the two application APIs.

\section{Discussion}
\label{sec:implications}

We investigated the feasibility of using documentation page-view logs to inform the design of documentation. 
Through a series of hypotheses derived from the literature, contextualized by an exploratory analysis of our page-view log data (\S\ref{sec:rq1}), and subsequently validated through a large-scale regression analysis (\S\ref{sec:regression}), we discovered that there are multiple discernible patterns of documentation use, even when the documentation pertains to the same platform, or even the same products.

\subsection{Feasibility of Log Analysis for Documentation Review}

\mysec{Large-scale log analysis helps discover unexpected use}
\revisiona{As large-scale log analysis allows analyzing all documentation usage, with less researcher efforts and costs, we could explore diverse documentation usage patterns.}
\revisiona{For example, in addition to users mainly using documentation for API learning, which was often studied in the existing literature that used smaller-scale qualitative approaches~\cite{meng2018application, meng2019developers, earle2015user, ko2006exploratory}, the clustering analysis discovered additional large clusters of users who}
only check pricing documentation (Cluster 22\revision{: financial users}), or that many users make many API requests without even visiting reference documentation (Cluster 27\revision{: task-oriented users}).
\revisiona{The cluster exploration and the hypotheses testing} also revealed that expected documentation usage can differ from actual use.
For example, although how-to documentation is often regarded as introductory for new users~\cite{earle2015user}, we observed that users with more product experience made more visits to the guide documentation (Cluster 6\revision{: task-oriented users}) than those with less product experience, which was also confirmed by the regression analyses (\hr{hyp:experience}).
While the cause or intent behind these unexpected uses cannot be found solely with log analysis, our observations might be useful in designing more focused human studies.
Moreover, we believe that a similar analysis can be used to
answer broader research questions like ``How does documentation usage change over time as users develop their expertise with the products?'', or ``What are the strategies developers use for information seeking in documentation?''

\mysec{Page-view log analysis is informative but could be further refined}
The analysis could be extended to also account for the structure and content of the documentation pages, in addition to the factors we considered.
For example, although the top web search results given the query \texttt{\revisionb{Google} [product]} were marketing documentation for all four products, the second result varied between a guide documentation page for \pubsub, and landing pages for \translate, \nl, and \logging.
Thus, in interpreting the differences in documentation usage between products, whether intended or not, differences between the documentation structure and external resources should also be taken into account.
Analyzing \textit{referrer} pages, \ie the pages accessed by a user prior to loading a particular web page, might be useful in understanding how such differences affect the documentation use~\cite{DBLP:journals/umuai/PierrakosPPS03}.
We propose this direction for future research.

\mysec{In practice, the analysis plan can be adapted based on the analysis goals}
In this paper, we employed a mixed-method approach to gain a comprehensive understanding of \company documentation usage. This involved both exploring the data and validating our hypotheses. Each of these analyses complements the other, offering distinct advantages and considerations.
For example, clustering analysis proves valuable in uncovering common and unexpected usage patterns, requiring less quantitative data analysis expertise to get started. However, it is important to note that interpreting clustering results can be subjective, and conducting a detailed investigation of every cluster may not always be practical.
Subsequently, performing regression analysis adds a layer of confidence to our findings, providing a comprehensive overview of the dataset.
In practice, it may not always be necessary or feasible to conduct both types of analysis due to differences in skill requirements. 
In such cases, the choice between the two can be made based on the specific goals of the log analysis.
For instance, a user experience (UX) researcher seeking a lightweight usability review might opt for a quick cluster analysis and interpretation as demonstrated in \Cref{subsec:sanity-test}. 
If stronger evidence is needed to support hypotheses, especially for design refactoring, engaging a quantitative UX researcher or data scientist to perform regression analysis following a clustering study would be a more suitable approach.

\subsection{Recommendations for Documentation Providers}
Through the log analysis, we found that documentation usage can vary based on the \revision{users' experience in product and platform} (\hr{hyp:experience}), the type of product described (\hr{hyp:product}), and many other factors (\hr{hyp:strategy}, \hr{hyp:intention}). 
This suggests that established knowledge on documentation usage may not always be generalizable to all target users.
Here, we highlight some of the specific implications for how to design improved documentation catering to users with different characteristics.

\mysec{Explicitly mention the target audience of documentation}\\
Previous studies~\cite{meng2018application} found that developers often experience difficulty in determining which documentation type to select when searching for a particular piece of information. 
We posit that this is because documentation for different products adheres to varying documentation standards and categorizes information differently, and it takes time for developers to learn these distinctions. 
Since we confirmed that developers' documentation visits are correlated with their characteristics, we posit that explicitly indicating the intended audience of the documentation will assist them in selecting the appropriate types and pages of documentation to access (\ie provide strong ``scent'' in the information foraging theory~\cite{10.1093/acprof:oso/9780195173321.001.0001}).

\mysec{{Duplicate important information for information discovery}}
{As our models show (\hr{hyp:strategy}), \revisionc{users are more likely to visit types of documentation when they have accessed in the past.} 
Although it is often considered to be better to \textit{modularize} the documentation, this can be problematic if important information is only presented on a specific page, as the user might not always discover that~\cite{horvath2019long,nam2023sorel}.
This observation is consistent with the finding of Meng et al. \cite{meng2018application} that developers often skip sections in the software documentation based on their problem solving strategies.
Thus, to reduce the risk of developers missing important information, we recommend providing such information in multiple types of documentation, or at least providing prominent functional links to the page providing such information.}

\mysec{Provide product-specific starting points}
We discovered that there are variations in visit patterns among products with distinct characteristics.
This is expected because different types of information are provided and needed depending on the purpose or domain of the product (\hr{hyp:product}).
For instance, for infrastructural products such as P3 and P4, many users (Cluster 2\revision{: task-oriented users}) accessed how-to guides providing instructions for the configuration settings, but for application products like P1 and P2, many users visited tutorial documentation pages providing walkthroughs for a simple use case (Cluster 21\revision{: product explorers}) that aid new users in quickly familiarizing themselves with the products.
However, for users who are new to the products with little understanding of them, it will be challenging to know what documentation type or page will be the best starting point~\cite{ko2007information}, especially because there are a plethora of documentation pages per product.
Thus, to help the new users quickly grasp the gist of the products, we recommend providing product-specific recommendations about which documentation pages to use to start learning, as similarly recommended in \citet{jeong2009improving}.
Most commonly accessed documentation pages or pages that correlate with subsequent API requests, which can be acquired from the page-view logs, will be good candidates for the recommendation, as they were already proven to be useful for other users.
We note that we do not recommend changing the documentation templates or navigation structures, because inconsistent inter-product information organization can hinder information foraging of users, especially those who use multiple products from \company.
A designated space for the product-specific documentation recommendation in a landing page or a navigation tab will allow users to know where to look if they become lost.

\mysec{Nudge new product users to visit guide-genre documentation}
When developers select third-party libraries, the quality of documentation is perceived as a good sign of the library's quality~\cite{DBLP:conf/sigsoft/VargasATBG20}, and when a user is not able to ﬁnd appropriate learning resources, it becomes a major obstacle in getting to know the libraries resulting in user frustration~\cite{10.1109/vlhcc.2016.7739689}.
Our results suggest that guide-genre documentation is particularly effective in influencing the decision to adopt a product (\hr{hyp:future}), although one might think that landing documentation is beneficial for them since it provides an overview of the products.
We believe that guide-genre documentation is helpful in making the adoption decision, as it describes what the products offer and help developers gauge what they need to do for onboarding, which corresponds to what new users look for from documentation~\cite{10.1109/vlhcc.2016.7739689}.
Thus, although other documentation pages will be useful in the end, nudging developers to visit guide-genre documentation as early as possible may help them perceive the quality of documentation positively, and adopt the API.

\subsection{Longer-term Vision: Personalization}

While we distilled actionable recommendations for how to adjust the design of software documentation taking into account many dimensions of user characteristics that might affect their usage, doing this manually may be unrealistic when many products are involved.
Instead, we argue that the time is ripe for approaches to \textit{automatically personalize} the documentation.
Personalization is not a new topic and has already proven to be effective for other services like media streaming and search engines~\cite{DBLP:journals/umuai/TintarevM12}.
Prior research on general web search has also made significant progress in designing effective personalized recommender systems to increase the long-term engagement of users~\cite{yi2014beyond}, using both implicit (\eg dwell time~\cite{yi2014beyond,DBLP:conf/kdd/YinLLW13}) and explicit (\eg item rating~\cite{DBLP:journals/ccsecis/AbdiOM18, DBLP:journals/cacm/AgarwalCER13}) feedback mined from historical interaction data as an indicator of users' interests and needs.
As the dwell time mined from documentation page-view logs can capture some user characteristics, in addition to the interaction histories that page-view logs contain by design, we expect that personalizing approaches can also be used in the documentation domain.
Here, we present three directions to improve developers' information foraging on documentation using page-view logs.

\mysec{Documentation recommendation}
First, we argue that it is time to go from static approaches of documentation recommendation (for example, consider the omnipresent navigation links like ``Recommended content'' or ``What’s next'' or ``Next topic,'' that typically point to the same target page regardless of which user is browsing) to dynamic ones that take user characteristics into account to provide more relevant suggestions. 
An ideal scenario is perhaps one where the recommender system has access to the developer’s code repository or profile, that reveal the developers' needs and background that are known to correlate with their documentation usage (\eg their product and platform experience), as we discovered from the analysis.
Short of that, we show that some signals about user-level characteristics are present in much more modest and more widely-available log data on previous documentation page visits. 
A recommender system could learn to profile users based on previous page visits (similar to our clustering) and, given that knowledge, suggest the next documentation pages to visit from among those that users in the same cluster have visited or interacted with before.

\mysec{Within-documentation search}
Personalization can also be applied to within-documentation search engines. 
Many previous studies of within-documentation search engines showed the need for efficient navigation~\cite{jeong2009improving}.
Typically, software documentation contains information for both novices and experts, sometimes implicitly within a single page, other times explicitly across dedicated separate pages. 
For example, a difference between a `basic' and an `advanced' tutorial could be that the advanced tutorial describes APIs with more flexible capabilities, which require additional parameters. 
One way to personalize 
is query modification~\cite{Shen.2005xq}, by expanding the user query using additional terms inferred from user profiles. 
As above, the user profiles can be approximated from documentation page view logs; for example, when a user’s documentation page view pattern is similar to Cluster 6 \revision{(task-oriented users)}, with high levels of guide documentation visits that correlate with product experience level, the system can infer that the user is experienced. 
Then, given a search query ``how to set up P1,'' the system could augment the query along the lines ``how to set up P1 \textit{advanced user},'' which should bias the search results towards the dedicated advanced pages.

\mysec{Documentation filtering}
Another idea is that a ``smart'' documentation system could automatically filter what information is being shown depending on the user. For example, when a user has already accessed platform-common information (e.g., authentication) from other products, the system can hide/fold such parts for new APIs the user is reading about, to make information foraging more efficient. 
Similarly, one could imagine hiding/folding other parts of a documentation page, such as the code examples, for users that prefer to develop a more conceptual understanding first~\cite{meng2018application}. 
These examples both require data on historical accesses of other documentation pages by the same users (or by users in the same cluster), which is often included in the page-view logs.

\section{Conclusion}

Based on our exploratory clustering analysis and hypothesis testing, we identified distinct documentation usage patterns and demonstrated that user factors partially explain the differences in such patterns. 
This enabled us to derive meaningful implications for documentation design, both specific to \company and in a broader context. 
Thus, we conclude that leveraging documentation logs at scale is both feasible and valuable and will allow documentation designers to generate actionable insights during their documentation design review.

\begin{acks} 
This research was funded in part by the NSF under grant CCF-2007482.
Any opinions, findings, and conclusions or recommendations expressed in this material are those of the authors and do not necessarily reflect those of the sponsors. 
We would like to thank anonymous external reviewers for their valuable feedback.
\end{acks}

\bibliographystyle{ACM-Reference-Format}
\bibliography{reference}
\balance

\clearpage
\pagebreak
\newpage
\appendix
\onecolumn

\setcounter{table}{0}
\renewcommand{\thetable}{A\arabic{table}}
\setcounter{figure}{0}
\renewcommand{\thefigure}{A\arabic{figure}}

\section{Dataset}
\begin{table*}[hbp!]
\caption{An example of our documentation page-view data used for the clustering analysis.}
\label{tab:vectors}
\resizebox{0.7\linewidth}{!}{
    \begin{tabular}{ccccccc}
    \toprule
    \multirow{2}{*}{User} & \multirow{2}{*}{Product}  & \multicolumn{5}{c}{Dwell Time (minutes)}   \\ \cmidrule{3-7} 
    &  & howto & marketing & reference & $\dots$ & concept \\ \midrule
    0 & \logging & 1     & 0  & 0       & $\dots$ & 0       \\ 
    1 & \logging & 35     & 0  & 0       & $\dots$ & 0       \\ 
    1 & \pubsub & 1.4  & 0    & 0       & $\dots$ & 0       \\ 
    $\vdots$    & $\vdots$    & $\vdots$    & $\vdots$   & $\vdots$ & $\vdots$       \\ 
    \bottomrule
    \end{tabular}
}
\end{table*}

\begin{table*}[hbp!]
\caption{An example of our API usage data used for the qualitative investigation of clustering results and the regression analysis.}
\label{tab:vectors}
\resizebox{0.7\linewidth}{!}{
    \begin{tabular}{ccccccc}
    \toprule
    User & Product & Account Age (years) & Past Succ. Req. & Future Succ. Req. \\ \midrule
    0 & \logging     & 5.96  & 0& 0       \\ 
    1 & \logging     & 2.33 & 1222& 859       \\ 
    1 & \pubsub      & 2.33    & 0& 0       \\ 
    $\vdots$           & $\vdots$& $\vdots$ & $\vdots$ &$\vdots$       \\ 
    \bottomrule
    \end{tabular}
}
\end{table*}

\begin{figure*}[hbp!]
  \centering
  \includegraphics[width=0.5\linewidth, trim={0 0 0 1cm}, clip]{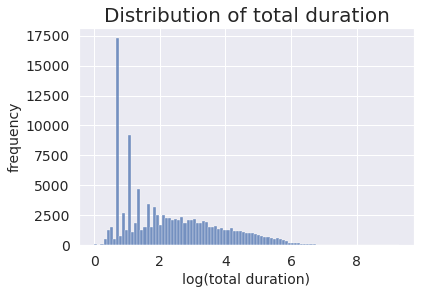}
  \caption{Distribution of the log-transformed total dwell time (in minutes) on documentation.}
  \label{fig:total_dwell_time}
\end{figure*}

% \vfill\null
\newpage

\section{Clustering analysis}

\begin{figure*}[hbp!]
  \centering
  \includegraphics[width=0.5\linewidth]{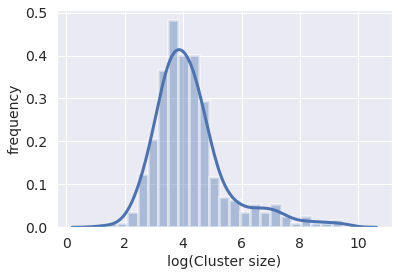}
  \caption{Distribution of log-transformed number of users per cluster.}
  \label{fig:cluster_size}
\end{figure*}

\clearpage

\begin{figure*}
\includegraphics[height=0.85\textheight, trim={2cm 3cm 3cm 4cm},clip]{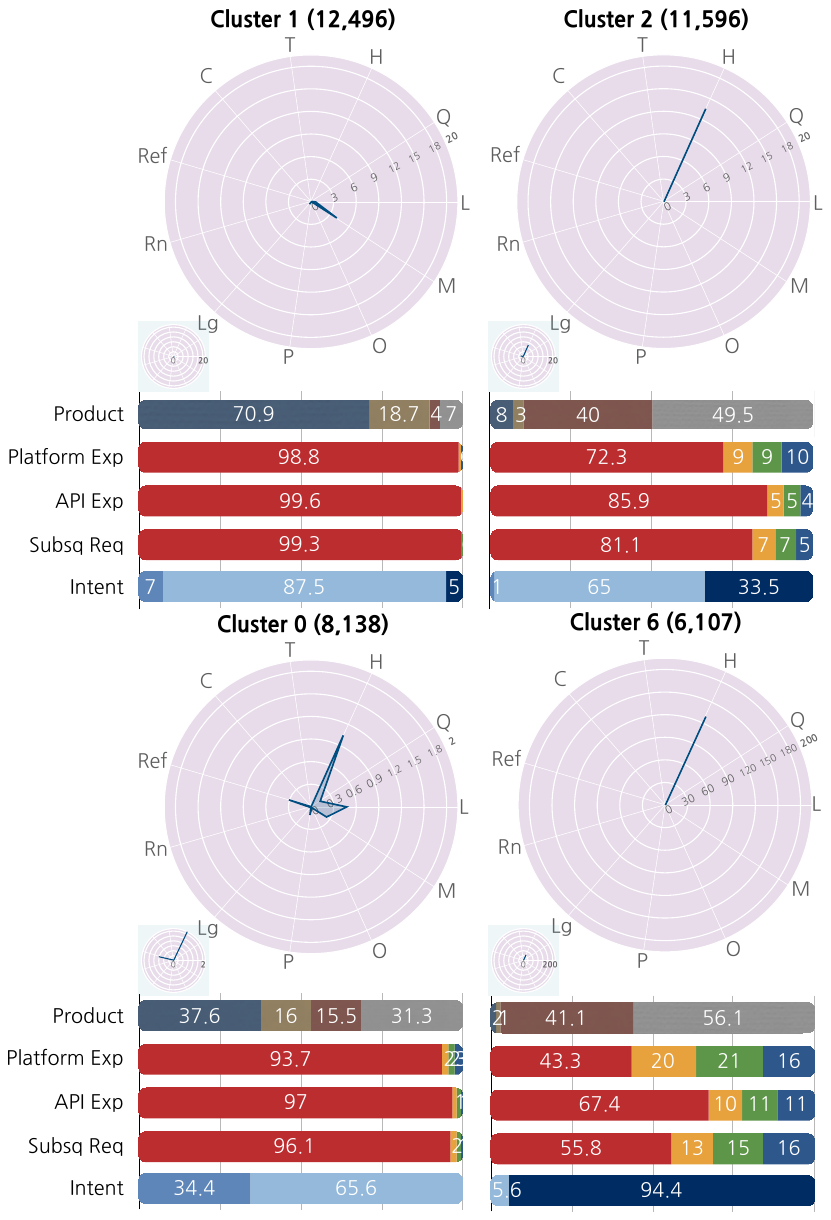}
\caption{Large clusters with more than 500 users sorted by the number of users (shown in parentheses).  Each polar plot displays the average time spent on each type of documentation (T: Tutorial, H: How-to, Q: Quickstart, L: Landing, M: Marketing, O: Other, P: Pricing, Lg: Legal, Rn: Release note, Ref: Reference, C:Concept). The small polar plots show the average dwell time in the previous three months. Note that the ranges of the axes of the plots vary. Bar charts below the polar plots show the proportions (\%) of each group in the cluster.}
\end{figure*}

\begin{figure*}
\includegraphics[height=0.85\textheight, trim={2cm 3cm 3cm 4cm},clip]{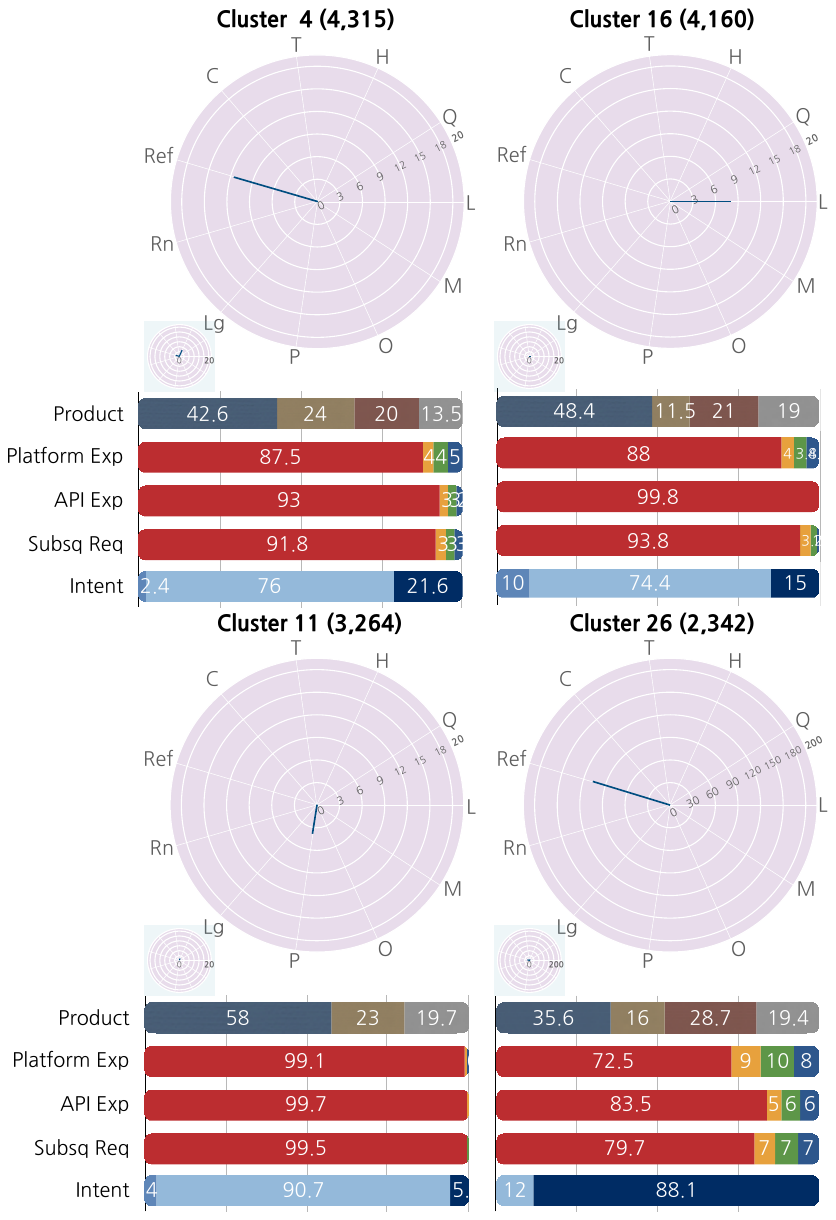}
\caption{Large clusters with more than 500 users sorted by the number of users (shown in parentheses).  Each polar plot displays the average time spent on each type of documentation (T: Tutorial, H: How-to, Q: Quickstart, L: Landing, M: Marketing, O: Other, P: Pricing, Lg: Legal, Rn: Release note, Ref: Reference, C:Concept). The small polar plots show the average dwell time in the previous three months. Note that the ranges of the axes of the plots vary. Bar charts below the polar plots show the proportions (\%) of each group in the cluster.}
\end{figure*}

\begin{figure*}
\includegraphics[height=0.85\textheight, trim={2cm 3cm 3cm 4cm},clip]{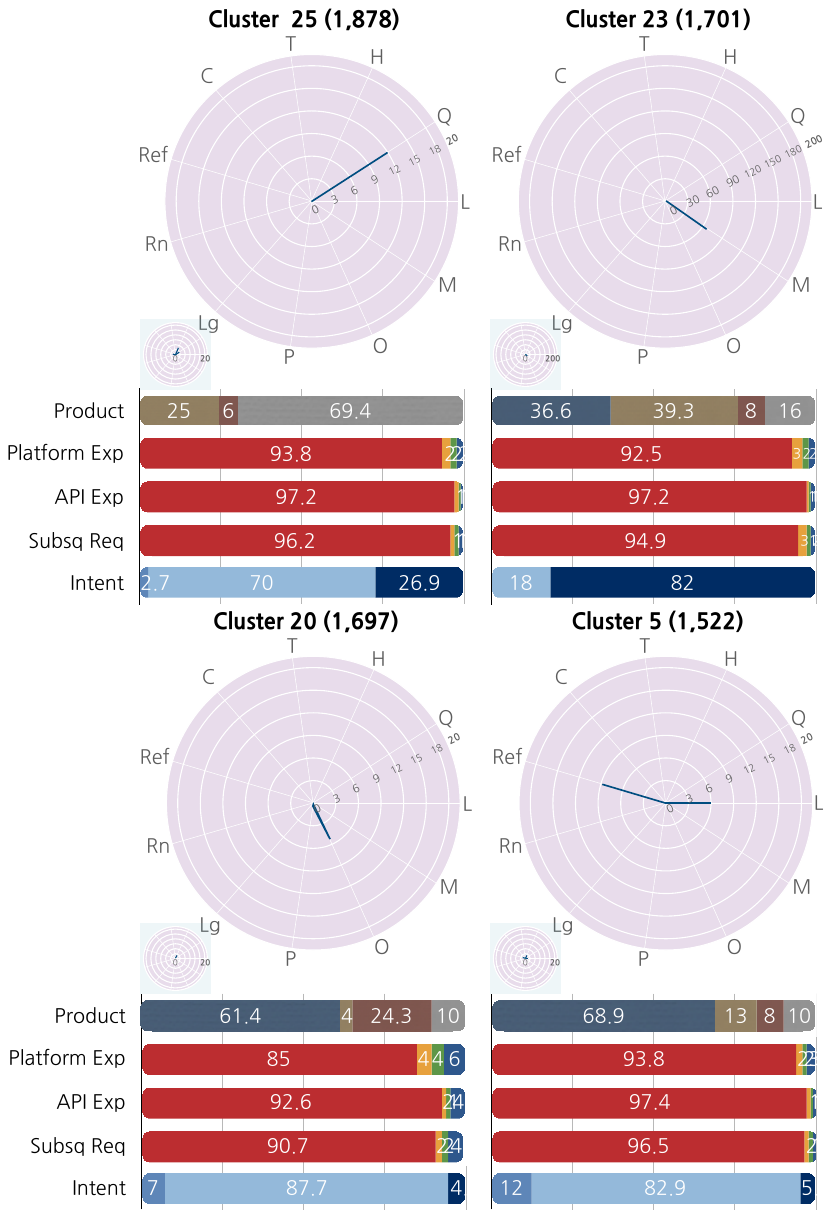}
\caption{Large clusters with more than 500 users sorted by the number of users (shown in parentheses).  Each polar plot displays the average time spent on each type of documentation (T: Tutorial, H: How-to, Q: Quickstart, L: Landing, M: Marketing, O: Other, P: Pricing, Lg: Legal, Rn: Release note, Ref: Reference, C:Concept). The small polar plots show the average dwell time in the previous three months. Note that the ranges of the axes of the plots vary. Bar charts below the polar plots show the proportions (\%) of each group in the cluster.}
\end{figure*}

\begin{figure*}
\includegraphics[height=0.85\textheight, trim={2cm 3cm 3cm 4cm},clip]{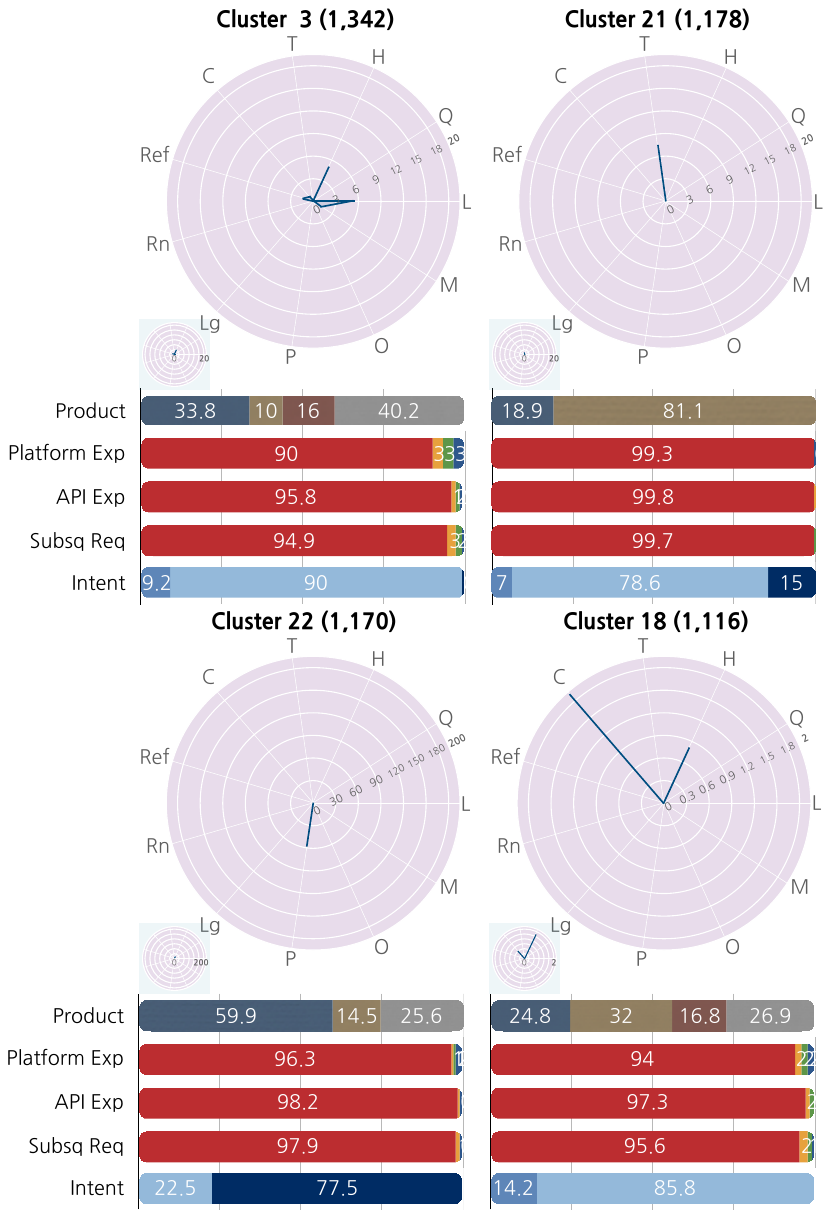}
\caption{Large clusters with more than 500 users sorted by the number of users (shown in parentheses).  Each polar plot displays the average time spent on each type of documentation (T: Tutorial, H: How-to, Q: Quickstart, L: Landing, M: Marketing, O: Other, P: Pricing, Lg: Legal, Rn: Release note, Ref: Reference, C:Concept). The small polar plots show the average dwell time in the previous three months. Note that the ranges of the axes of the plots vary. Bar charts below the polar plots show the proportions (\%) of each group in the cluster.}
\end{figure*}

\begin{figure*}
\includegraphics[height=0.85\textheight, trim={2cm 3cm 3cm 4cm},clip]{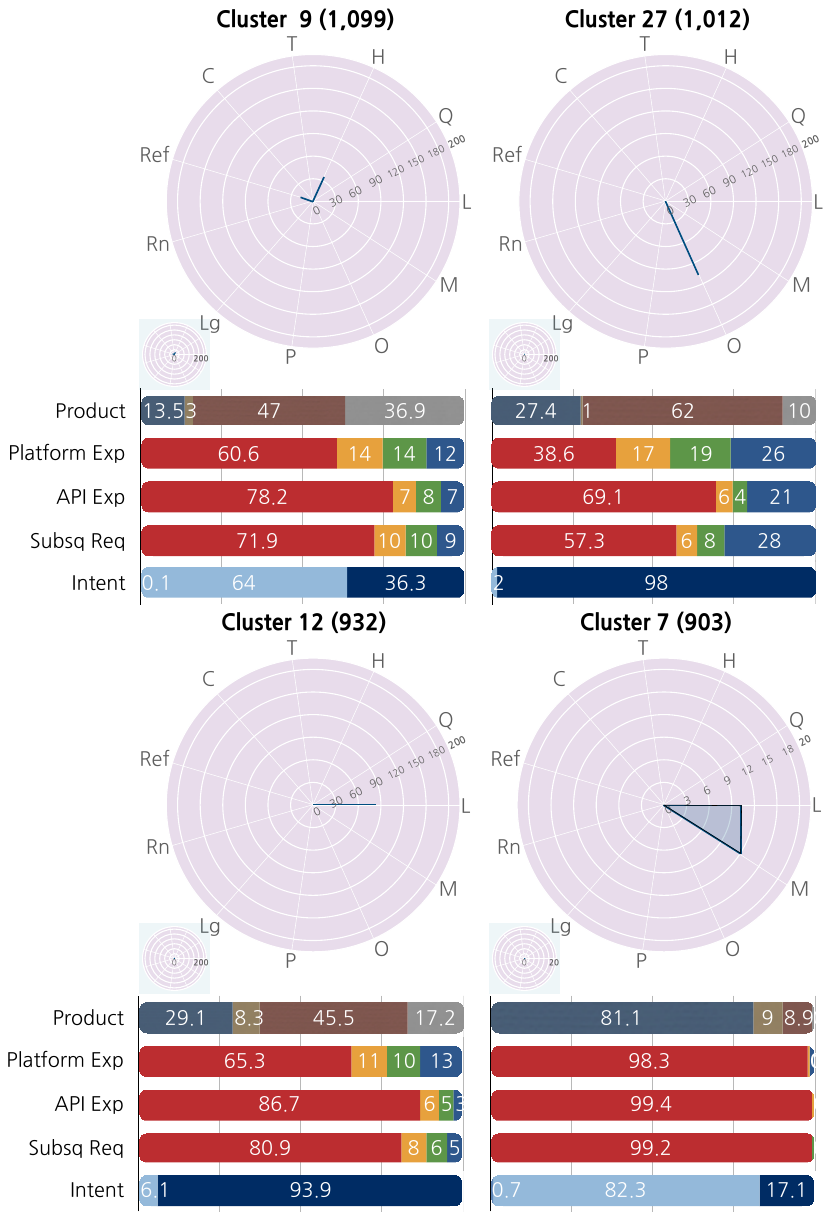}
\caption{Large clusters with more than 500 users sorted by the number of users (shown in parentheses).  Each polar plot displays the average time spent on each type of documentation (T: Tutorial, H: How-to, Q: Quickstart, L: Landing, M: Marketing, O: Other, P: Pricing, Lg: Legal, Rn: Release note, Ref: Reference, C:Concept). The small polar plots show the average dwell time in the previous three months. Note that the ranges of the axes of the plots vary. Bar charts below the polar plots show the proportions (\%) of each group in the cluster.}
\end{figure*}

\begin{figure*}
\includegraphics[height=0.85\textheight, trim={2cm 3cm 3cm 4cm},clip]{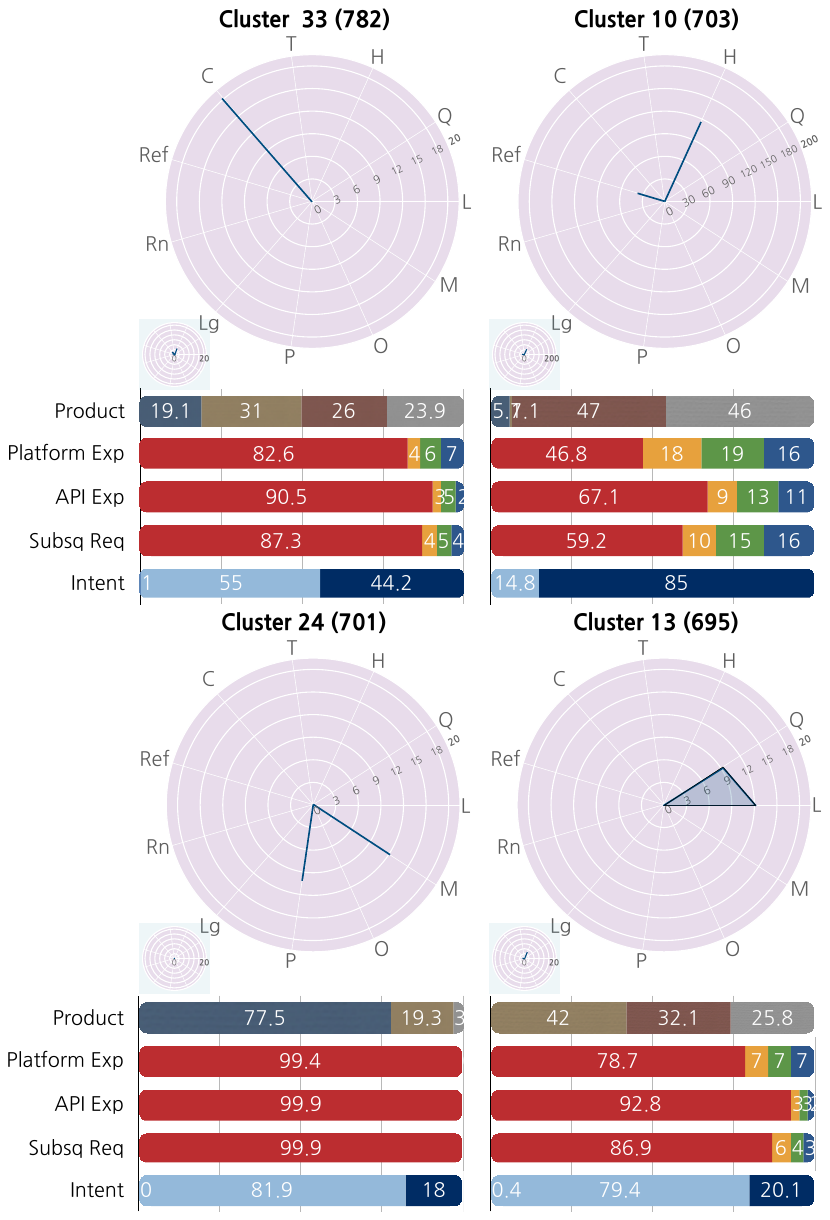}
\caption{Large clusters with more than 500 users sorted by the number of users (shown in parentheses).  Each polar plot displays the average time spent on each type of documentation (T: Tutorial, H: How-to, Q: Quickstart, L: Landing, M: Marketing, O: Other, P: Pricing, Lg: Legal, Rn: Release note, Ref: Reference, C:Concept). The small polar plots show the average dwell time in the previous three months. Note that the ranges of the axes of the plots vary. Bar charts below the polar plots show the proportions (\%) of each group in the cluster.}
\end{figure*}

\begin{figure*}
\includegraphics[height=0.85\textheight, trim={2cm 3cm 3cm 4cm},clip]{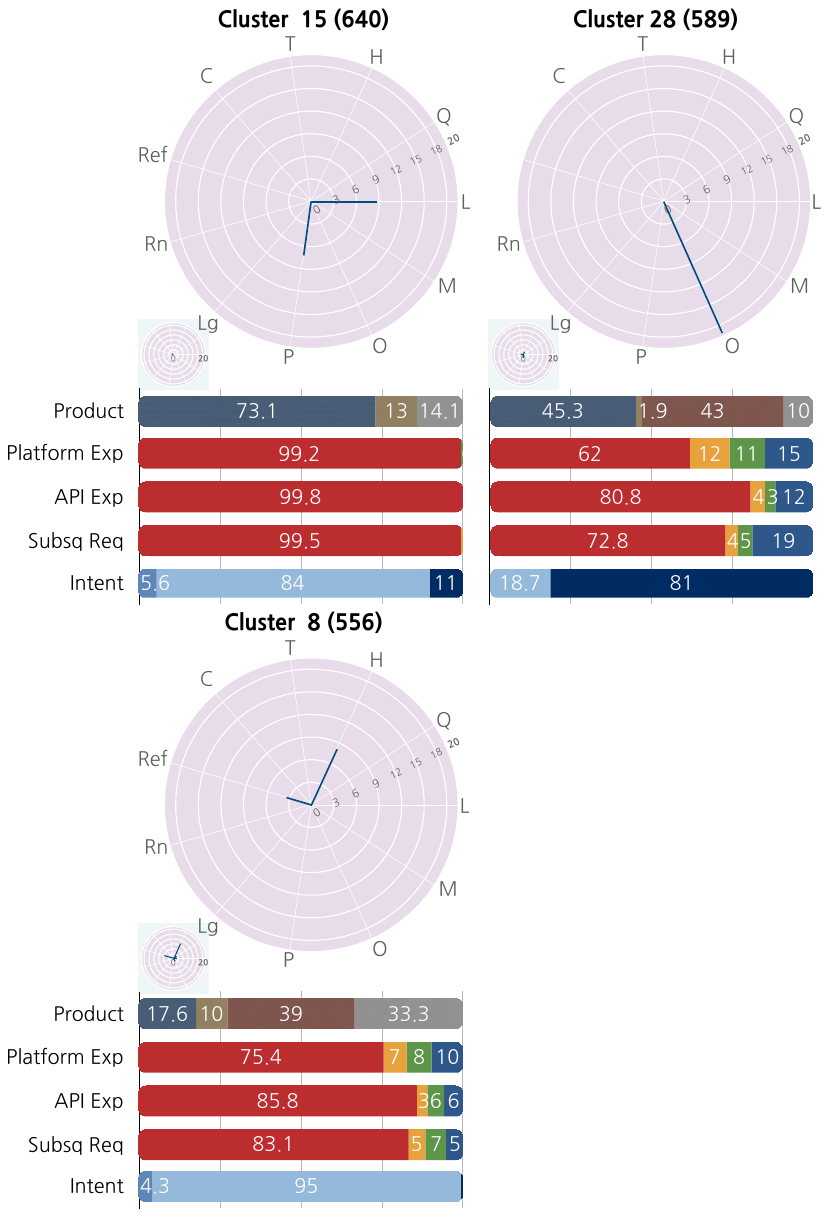}
\caption{Large clusters with more than 500 users sorted by the number of users (shown in parentheses).  Each polar plot displays the average time spent on each type of documentation (T: Tutorial, H: How-to, Q: Quickstart, L: Landing, M: Marketing, O: Other, P: Pricing, Lg: Legal, Rn: Release note, Ref: Reference, C:Concept). The small polar plots show the average dwell time in the previous three months. Note that the ranges of the axes of the plots vary. Bar charts below the polar plots show the proportions (\%) of each group in the cluster.}
\end{figure*}

\clearpage

\section{Regression analysis}

Example regression model specification for the guide documentation genre: 

{\small
$$
\begin{aligned}[hbp!]
\log&\left[ \frac { P( \text{guide\_dwell\_time} > 0 ) }{ 1 - P( \text{guide\_dwell\_time} > 0 ) } \right] = \alpha +\\ 
&\beta_{1}(\text{overall\_platform\_experience}) +\\ &\beta_{2}(\text{average\_page\_dwell\_time}) +\\ 
&\beta_{3}(\text{product}) +\\
&\beta_{4}(\text{specific\_API\_experience} > 0) +\\
&\beta_{5}(\text{dev\_page\_views\_in\_the\_previous\_three\_months} > 0) +\\
&\beta_{6}(\text{guide\_page\_views\_in\_the\_previous\_three\_months} > 0) +\\
&\beta_{7}(\text{admin\_page\_views\_in\_the\_previous\_three\_months} > 0) +\\
&\beta_{8}(\text{meta\_page\_views\_in\_the\_previous\_three\_months} > 0)
\end{aligned}
$$
}
\newpage

\begin{figure*}[hbp!]
    \centering
    \includegraphics[width=0.6\linewidth]{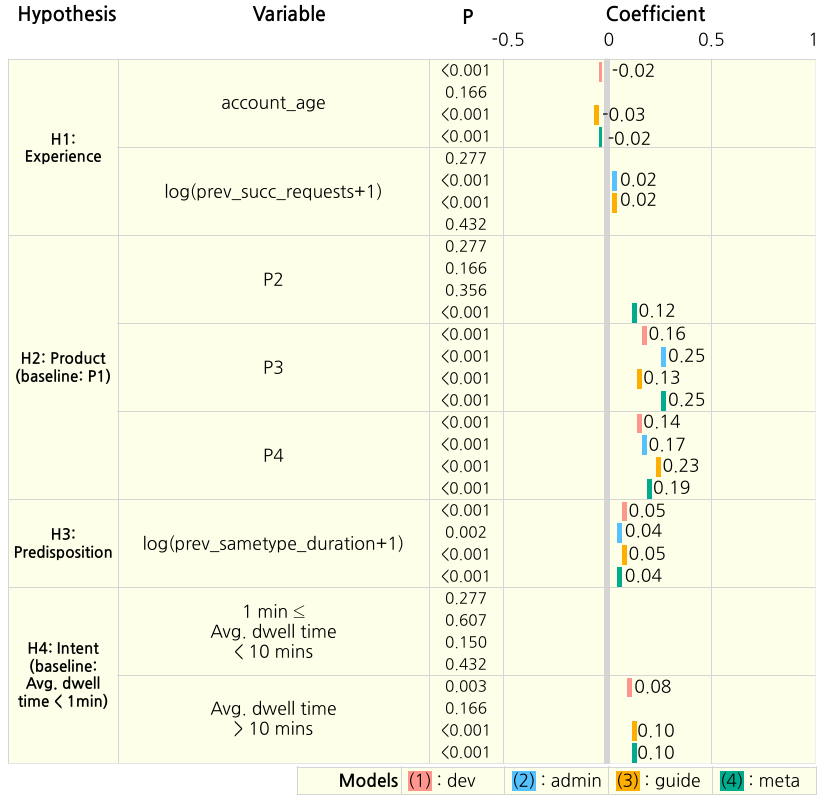}
    \caption{Coefficients from regression analysis predicting dwell time for four types of documentation. p-values are adjusted based on the Holm's correction~\cite{holm1979simple}. Coefficients are removed for non-significant results (p\textgreater.001).}
    \label{fig:rq2-linear}
\end{figure*}

\begin{figure*}[hbp!]
    \centering
    \includegraphics[width=0.6\linewidth]{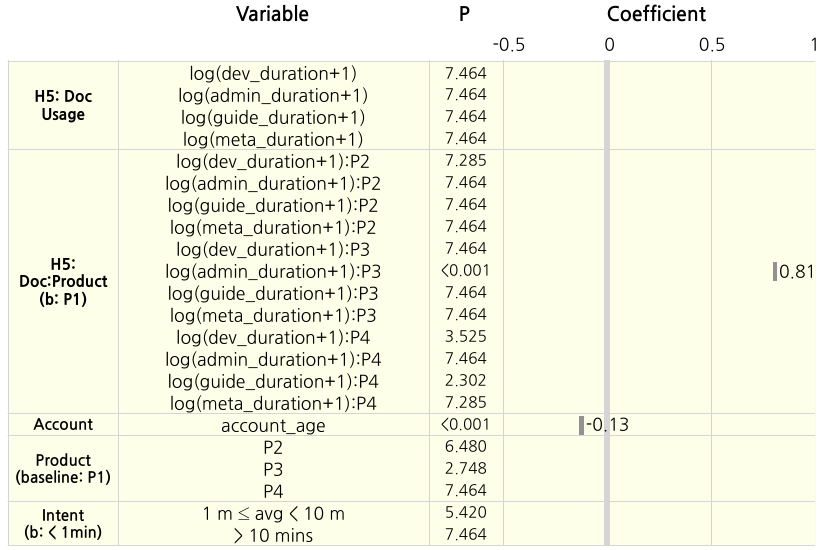}
    \caption{Coefficients from regression analysis predicting dwell time for four types of documentation. p-values are adjusted based on the Holm's correction~\cite{holm1979simple}. Coefficients are removed for non-significant results (p\textgreater.001).}
    \label{fig:rq3-linear}
\end{figure*}

\end{document}